\documentclass[useAMS,usenatbib,usegraphicx]{mn2e}

\usepackage{amssymb}
\usepackage{wasysym}
\usepackage{multirow}
\usepackage[usenames]{color}
\DeclareSymbolFont{boldss}{OT1}{cmss}{bx}{n}
\DeclareSymbolFontAlphabet\mathbsf{boldss}

\def\jref@jnl#1{{\rm#1}}

\def\aj{\jref@jnl{AJ}}                   
\def\araa{\jref@jnl{ARA\&A}}             
\def\apj{\jref@jnl{ApJ}}                 
\def\apjl{\jref@jnl{ApJ}}                
\def\apjs{\jref@jnl{ApJS}}               
\def\ao{\jref@jnl{Appl.~Opt.}}           
\def\apss{\jref@jnl{Ap\&SS}}             
\def\aap{\jref@jnl{A\&A}}                
\def\aapr{\jref@jnl{A\&A~Rev.}}          
\def\aaps{\jref@jnl{A\&AS}}              
\def\azh{\jref@jnl{AZh}}                 
\def\baas{\jref@jnl{BAAS}}               
\def\jrasc{\jref@jnl{JRASC}}             
\def\memras{\jref@jnl{MmRAS}}            
\def\mnras{\jref@jnl{MNRAS}}             
\def\pra{\jref@jnl{Phys.~Rev.~A}}        
\def\prb{\jref@jnl{Phys.~Rev.~B}}        
\def\prc{\jref@jnl{Phys.~Rev.~C}}        
\def\prd{\jref@jnl{Phys.~Rev.~D}}        
\def\pre{\jref@jnl{Phys.~Rev.~E}}        
\def\prl{\jref@jnl{Phys.~Rev.~Lett.}}    
\def\pasp{\jref@jnl{PASP}}               
\def\pasj{\jref@jnl{PASJ}}               
\def\qjras{\jref@jnl{QJRAS}}             
\def\skytel{\jref@jnl{S\&T}}             
\def\solphys{\jref@jnl{Sol.~Phys.}}      
\def\sovast{\jref@jnl{Soviet~Ast.}}      
\def\ssr{\jref@jnl{Space~Sci.~Rev.}}     
\def\zap{\jref@jnl{ZAp}}                 
\def\nat{\jref@jnl{Nature}}              
\def\iaucirc{\jref@jnl{IAU~Circ.}}       
\def\aplett{\jref@jnl{Astrophys.~Lett.}} 
\def\apspr{\jref@jnl{Astrophys.~Space~Phys.~Res.}}
\def\bain{\jref@jnl{Bull.~Astron.~Inst.~Netherlands}} 
\def\fcp{\jref@jnl{Fund.~Cosmic~Phys.}}  
\def\gca{\jref@jnl{Geochim.~Cosmochim.~Acta}}   
\def\grl{\jref@jnl{Geophys.~Res.~Lett.}} 
\def\jcp{\jref@jnl{J.~Chem.~Phys.}}      
\def\jgr{\jref@jnl{J.~Geophys.~Res.}}    
\def\jqsrt{\jref@jnl{J.~Quant.~Spec.~Radiat.~Transf.}}
\def\memsai{\jref@jnl{Mem.~Soc.~Astron.~Italiana}}
\def\nphysa{\jref@jnl{Nucl.~Phys.~A}}   
\def\physrep{\jref@jnl{Phys.~Rep.}}   
\def\physscr{\jref@jnl{Phys.~Scr}}   
\def\planss{\jref@jnl{Planet.~Space~Sci.}}   
\def\procspie{\jref@jnl{Proc.~SPIE}}   

\def\T1{{\rm 1T}}

\title[Polytropic dark halos of elliptical galaxies]{
   Polytropic dark halos of elliptical galaxies}
\author[C.~J.~Saxton \& I.~Ferreras]
{ Curtis~J.~Saxton\thanks{E-mail: cjs2@mssl.ucl.ac.uk} and Ignacio Ferreras\\
Mullard Space Science Laboratory, University College London, 
   Holmbury St Mary, Dorking, Surrey RH5 6NT  \\
}

\date{MNRAS Ref  MN-09-2224-MJ, Revised version - January 2010}
\pagerange{\pageref{firstpage}--\pageref{lastpage}} \pubyear{2010}
\begin{document}


\maketitle
\label{firstpage}
\begin{abstract}
The kinematics of stars and planetary nebulae in early type galaxies
provide vital clues to the enigmatic physics of their dark matter
halos.  We fit published data for fourteen such galaxies using a
spherical, self-gravitating model with two components: (1) a S\'ersic
stellar profile fixed according to photometric parameters, and (2) a
polytropic dark matter halo that conforms consistently to the shared
gravitational potential.  The polytropic equation of state can
describe extended theories of dark matter involving self-interaction,
non-extensive thermostatistics, or boson condensation (in a classical
limit).  In such models, the flat-cored mass profiles widely observed
in disc galaxies are due to innate dark physics, regardless of any
baryonic agitation.  One of the natural parameters of this scenario is
the number of effective thermal degrees of freedom of dark matter
($F_{\rm d}$) which is proportional to the dark heat capacity.  By
default we assume a cosmic ratio of baryonic and dark mass.
Non-S\'ersic kinematic ideosyncrasies and possible non-sphericity
thwart fitting in some cases. In all fourteen galaxies the fit with a
polytropic dark halo improves or at least gives similar fits to the
velocity dispersion profile, compared to a stars-only model. The good
halo fits usually prefer $F_{\rm d}$ values from six to eight.  This
range complements the recently inferred limit of $7<$$F_{\rm d}<10$
(Saxton \& Wu), derived from constraints on galaxy cluster core radii
and black hole masses.  However a degeneracy remains: radial orbital
anisotropy or a depleted dark mass fraction could shift our models'
preference towards lower $F_{\rm d}$; whereas a loss of baryons would
favour higher $F_{\rm d}$.
\end{abstract}


\section{Introduction}
Simulations that assume simple, non-interacting dark matter
successfully reproduce cosmic large-scale structure \citep[see
e.g.][]{sfw06}. However, they also predict power-law cusps of dark
density at the centres of all virialised structures
\citep[e.g.][]{dubinski1991,nfw1996,moore1998,diemand2004,diemand2005}.
Dynamical and other circumstantial evidence for many galaxy types
imply cores of nearly uniform dark density
\citep[e.g][]{flores1994,moore1994,burkert1995,weldrake2003,
gentile2004,simon2005,deblok2005,kuzio2006,weijmans2008}.  It remains
debatable whether the present-day cores are innate, or whether a
violent, primordial form of baryonic ``feedback'' somehow conspired to
erase (or exacerbate) all cusps
\citep[e.g.][]{blumenthal1986,gnedin2002,gnedin2004,
romano-diaz2008,jardel2008}.

The observational situation in elliptical galaxies is less clear than
for other morphological types, because of the relative difficulty of
observing suitable kinematic tracers at a range of radii.  These data
may suffer degeneracies between mass profiles and the orbital
anisotropies of kinematic tracers \citep{binney82}.  X-ray emission
(if detected) is informative if hydrostasis is a fair assumption.
Observations of some X-ray emitting elliptical galaxies favour the
presence of massive dark halos
\citep[e.g.][]{awaki1994,loewenstein1999,mathews2003,humphrey2006}.
For some controversial galaxies, dark matter appears dynamically
absent, perhaps more diffuse than in standard cosmogonies, or possibly
disguised by stellar orbital anisotropies
\citep[e.g.][]{mendez2001,romanowsky2003,dekel2005,delorenzi2008b}.
In other cases the inference of dark matter is clearer
\citep[e.g][]{osullivan2004,delorenzi2008a}. Gravitational lens models
suggest steep, isothermal slopes around the observed radii.  However,
inner regions show moderate dark matter densities
\citep[e.g.][]{gerhard2001,ferreras2005,cappellari2006,thomas2007,weijmans2009},
despite the cuspy predictions of simple collisionless dark matter
cosmogonies, and the expectation of adiabatic contraction due to the
infall of the baryons towards the centre of the halo
\citep[e.g.][]{blumenthal1986}.

Before N-body simulations became popular and highly resolved,
flat-cored profiles like the ``pseudo-isothermal sphere'' were by
default assumed in studies of galaxies and galaxy clusters.
Observations of galaxy-scale dark cores now motivate the examination
of alternative dark matter theories where the physics naturally cause
cores, regardless of any baryonic ``feedback''
\citep[e.g.][]{spergel2000,peebles2000,arbey2003,ahn2005,saxton2008}.
\citet{nunez2006} and \citet{zavala2006} compared the scaling
relations of disc galaxies using either a standard dark matter profile
\citep[NFW,][]{nfw1996}, or a polytrope -- intended to represent
collisionless dark matter subject to Tsallis thermostatistics
\citep{tsallis1988} -- and found the latter to explain better the
observations.
\citet{boehmer2007} also found good fits to the rotation curves
of low surface brightness galaxies and dwarf galaxies with a
polytropic dark matter halo (which they justify via the formation of a
Bose-Einstein condensate). 

Furthermore, polytropic dark matter halos can equally represent other
interesting scenarios, such as self-interacting dark matter, where the
core has formed due to the support of dark pressure in the dark field
domain.  Polytropic models are naturally consistent, self-gravitating
distributions of classical mass, and may be seen as analogues to
adiabatic fluids.  Some early simulations with numerical
approximations to dark hydrodynamics
\citep{moore2000,yoshida2000b,burkert2000,kochanek2000} entailed
(dark) thermal conduction leading to gravothermal catastrophe, and
hence isothermal cusps that were sharper than those of collisionless
simulations. However it was eventually shown
\citep{balberg2002a,balberg2002b,ahn2005} that the simulations were
unwittingly (and inappropriately) initiated in near-collapse
conditions.  It was shown that conductivity is only significant for
SIDM of medium interactivity.  Weaker SIDM has infrequent collisions
and conduction is slow.  Strongly self-interactive dark matter has
vanishing conductivity (as it returns to an adiabatic limit) which
defers collapse well beyond the age of the universe.

In this paper we extend the polytropic dark halo model of
\citet{saxton2008} to galaxy scales. We apply the model to fitting
planetary nebula and stellar kinematics of elliptical galaxies. We
present fits for a number of cases, including ``stars only'' models,
``stars + halo'' models with isotropic orbits and a constant global
baryon fraction, and finally we also consider the issue of orbital
anisotropy and a non-universal baryon fraction.

\section{Method}

\subsection{Spherical model and its solutions}
We assume a fixed stellar mass distribution, with a projected surface
density approximately described by the \citet{sersic1968} profile.
S\'ersic-like profiles appear ubiquitous among observed galaxy
spheroids, although the reasons why this form emerges in nature have
not yet been demonstrated analytically.  For the three-dimensional
stellar density, we adopt the \citet{prugniel1997} spherical
deprojection \citep[see also][]{limaneto1999,cotti1999,marquez2000}.
\begin{equation}
	\rho_\bigstar(x) = \rho_{\rm e} x^{-p}
	\exp\left[{-b\left({x^{1/n}-1}\right)}\right]
	\ ,
\end{equation}
   with dimensionless coordinate $x\equiv r/r_{\rm e}$,
   and $(b,p)$ are constants depending on the index $n$.
The scale radius (3D) is approximately equal to
   the observational, projected (2D) half-light radius,
   $r_{\rm e}\approx R_{\rm e}$.
The stellar mass within some radius is given by
\begin{equation}
	m_\bigstar(<x) = 4\pi n b^{n(p-3)} e^b
	\rho_{\rm e} r_{\rm e}^3
	\ \Gamma[n(3-p),b x^{1/n}]
\end{equation}
involving lower incomplete gamma functions.

Within this empirical stellar distribution, we introduce a
dark matter halo, with a density distribution to be solved
consistently with the global gravitational potential.  The local
equation of state is polytropic,
\begin{equation}
	P_{\rm d} = s_{\rm d}\,\rho_{\rm d}^{\gamma_{\rm d}}
\end{equation}
with a pressure $P_{\rm d}$, density $\rho_{\rm d}$, pseudo-entropy
$s_{\rm d}$ and adiabatic index $\gamma_{\rm d}\equiv 1 + 2/F_{\rm d}$.  
This is the natural behaviour expected for an isotropic medium
comprising classical particles with $F_{\rm d}$ effective degrees of freedom.
Such a condition is expected if dark matter is self-interacting
\citep[e.g.][]{spergel2000,peebles2000,arbey2003,ahn2005,saxton2008}.
An equivalent polytropic equation emerges if the dark halo is governed
by Tsallis' non-extensive thermostatistics
\citep[e.g.][]{tsallis1988,plastino1993,zavala2006,nunez2006}.

If 
$-2<F_{\rm d}\le10$
then the dark halo may truncate at finite mass and
radius ($R_{\rm h}$).  For any particular model, we find this radius by
assuming dark hydrostasis
\begin{equation}
	{{dP_{\rm d}}\over{dr}} =
	-{{G\,m\,\rho_{\rm d}}\over{r^2}}
	\ ,
\end{equation}
where we define the interior mass $m=m_\bigstar(r)+m_{\rm d}(r)$ for
brevity.  Computationally, it is more convenient to deal with the dark
velocity dispersion (or temperature) 
$\sigma^2_{\rm d}=P_{\rm d}/\rho_{\rm d}$, which obeys
\begin{equation}
	{{d\sigma^2_{\rm d}}\over{dr}} =
	-\left({2\over{F_{\rm d}+2}}\right)
	{{G\,m}\over{r^2}}
	\ .
\end{equation}
We take $s_{\rm d}$ as global constant, and numerically integrate the
dark profile from the origin outwards.  We assume non-singular
conditions at the origin,
\begin{equation}
	\left.{{d\rho_{\rm d}}\over{dr}}\right|_{r\rightarrow0}=0
	\ ,
\end{equation}
precluding the complication of a compact central mass and its sphere
of influence.  At some point the dark velocity dispersion becomes small
(also $\rho_{\rm d}\rightarrow0$ if $F_{\rm d}>0$)
and the
integrator switches to use $\sigma^2_{\rm d}$ as the independent
variable instead of $r$.  Integration proceeds exactly to the point
where $\sigma^2_{\rm d}=0$, i.e. the natural truncation radius $R_{\rm h}$.
We call this the ``dark surface''.
The low-mass region between the dark core
(say where the log-slope of density is $-2$)
and the dark surface shall be called the ``halo fringe.''

Once the total dark mass $M_{\rm d}$ and $R_{\rm h}$ are known, we can obtain
the external gravitational potential, stellar pressure, potential
energy and thermal energy profiles by integrating a Jeans equation
(and associated ODEs for the mass profile and potential) inwards from
infinity.  If we define a stellar pressure in terms of the radial
velocity dispersion of stars,
$P_\bigstar\equiv\rho_\bigstar\sigma^2_\bigstar$ then we have the
spherical Jeans equation

\begin{equation}
	{{dP_\bigstar}\over{dr}}
	= -{{G\,m\,\rho_\bigstar}\over{r^2}} -{{2\beta P_\bigstar}\over{r}}
\label{eq.jeans}
\end{equation}
where $\beta\equiv1-\sigma_\theta^2/\sigma^2_\bigstar$ is the local
stellar velocity anisotropy.  When integrating (\ref{eq.jeans}) far
outside the dark halo, we change the independent variable to
$u\equiv1/r$.  Having obtained all the relevant constants of
integration at the dark surface $R_{\rm h}$, we can easily evaluate profiles
of the stellar and dark components in both the interior and exterior.

\subsection{Model specification and projection}

In any theory with $F_{\rm d}$ dark matter degrees of freedom fixed,
the pseudo-entropy $s_{\rm d}$ and central dark temperature are
formally free parameters.  When discussing or fitting families of
physically related models, it is more useful to prescribe the stellar
mass fraction within some radius, $\mu_\bigstar(r) =
m_\bigstar/(m_\bigstar+m_{\rm d})$, or the dark fraction 
$\mu_{\rm d}=1-\mu_\bigstar$.  If galaxies retain all their primordial
endowments of matter and darkness, then total stellar value would
approximate the cosmic baryon fraction
$\mu_\bigstar(\infty)\approx0.16$\footnote{We note that the latest
WMAP-5 results  give a best fit for a cosmological
baryon fraction of $\sim$0.2 \citep{wmap5}. We refer the reader to 
\S\S3.4 for an estimate of a change in this number.}
When seeking particular numerical
solutions, we shall attain a chosen $\mu_\bigstar(\infty)$ by fixing
the dark entropy $s_{\rm d}$ and adjusting the central $\sigma^2_{\rm
d}$ iteratively.  Then (for a given stellar background) there is an
implicit relation between $s_{\rm d}$ and the dark truncation radius
$R_{\rm h}$.

Model observables are calculated by integrating suitable 3D stellar
properties at radii $r=\sqrt{z^2+b^2}$ where $b$ is the projected
radius on the sky plane and $z$ is the line-of-sight coordinate.  The
column density of stars is simply

\begin{equation}
	\Sigma_\bigstar = 2\,\int_0^\infty \rho_\bigstar\,dz
	\ ,
\label{eq.column}
\end{equation}
   and the sightline integrated pressure is
\begin{equation}
	\Pi_\bigstar = 2\,\int_0^\infty P_\bigstar
		\left[{(1-\beta)+\beta{{z^2}\over{r^2}} }\right]\,dz
	\ .
\label{eq.starpress}
\end{equation}
The mean line-of-sight velocity dispersion is then $\sigma_{\rm
los}=\sqrt{\Pi_\bigstar/\Sigma_\bigstar}$.  The orbital anisotropy is
crucial to the stellar observables.  We will mainly consider models
with perfect isotropy ($\beta=0$ everywhere).  In later sections
(\S\ref{s.anisotropy}),
we will consider anisotropy varying radially according to the
Osipkov--Merritt model \citep{osipkov1979,merritt1985},
\begin{equation}
	\beta(r) = {{r^2}\over{r^2 + a^2}}
\label{eq.OM}
\end{equation}
with the scale radius $a$ treated as a fitting parameter.

In practice, integration of the stellar profile to radii at large
orders of magnitude eventually samples a tenuous, hot region where
$\rho_\bigstar$ drops much faster than $P_\bigstar$.  Formally, the
integrated velocity dispersion rises greatly.  Physically, this
contribution should not occur, because the densities imply fewer than
one star within the relevant volume.  Therefore we cut off the
integrals (\ref{eq.column}) and (\ref{eq.starpress}) at some
three-dimensional radius far outside the kinematic tracers of a
particular galaxy.

Calculations are performed in natural units of the S\'ersic profile:
$\rho_{\rm e}=R_{\rm e}=G=1$.  A numerical minimal-$\chi^2$ fitting
procedure automatically yields the normalisation for a particular data
set. This auto-normalisation counts as an extra free parameter,
decrementing the ``degrees of freedom'' in the reduced $\chi^2$
scores.

\section{Results}

We fit the kinematics of stars and planetary nebulae of a sample
comprising twelve galaxies from \citet{coccato2009} as well as NGC~1400
and NGC~1407.  These data are separated into major and minor axis
components.  We fit the axes separately, to obtain an indication of
the applicability of our model's spherical assumption.  Additionally,
we fit kinematic data for two of the galaxies without decomposition
along axes: for NGC~821 we extract data from 
\citet{romanowsky2003}.  For NGC~3379 we use figure~11 from
\citet{douglas2007}. The data for NGC~1400 and 1407 are taken 
from \citet{proctor2009}. Notice in this case the data are extracted
over a circular aperture, so only one axis can be considered.

We assume that the stellar mass component follows the profile of
\cite{prugniel1997}, with the S\'ersic parameters shown in the left
three columns of Table~\ref{table.stars.optima}.  For nine of the
galaxies, these parameters come from photometric fits to S\'ersic-only
models by \citet[][their table 3]{coccato2009}.  For NGC~3379 we take
$n=4.7$ from \citet{douglas2007}.  For NGC~5128 we assume $n=4$ and
obtain $R_{\rm e}$ from the B-band photometric fit of
\citet{dufour1979}.  For NGC~5846, we convert S\'ersic parameters from
\citet{mahdavi2005b}.  For all galaxies, we keep the distances and
luminosities tabulated by \citet{coccato2009}.  Initially we attempt
fits without a polytrope (``stars only'') to test the necessity of
dark matter, and to provide benchmarks for the fitting qualities of
halo models.

\subsection{Fits with stars only}

\begin{figure*}
\begin{center} 
\begin{tabular}{ccc}
 \includegraphics[width=7cm]{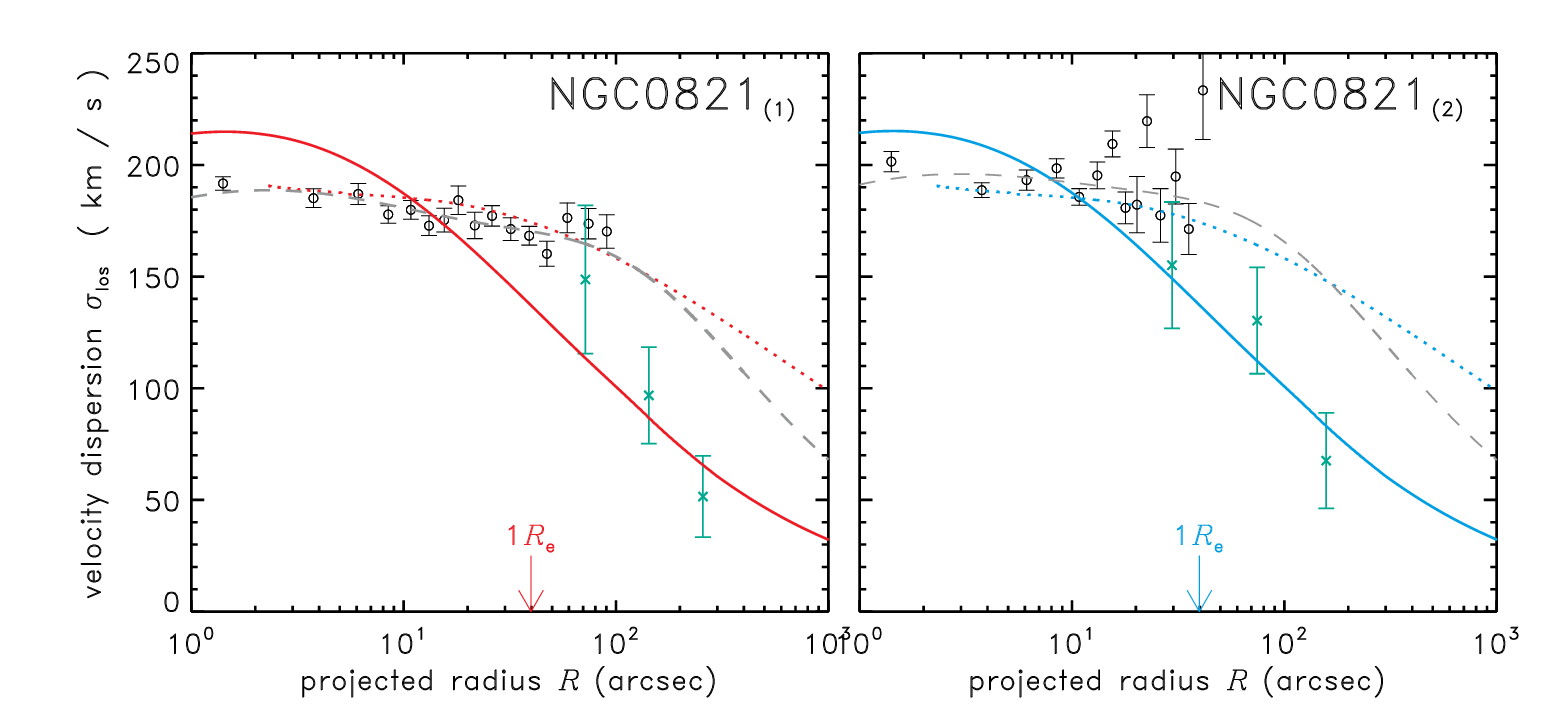}
&\includegraphics[width=7cm]{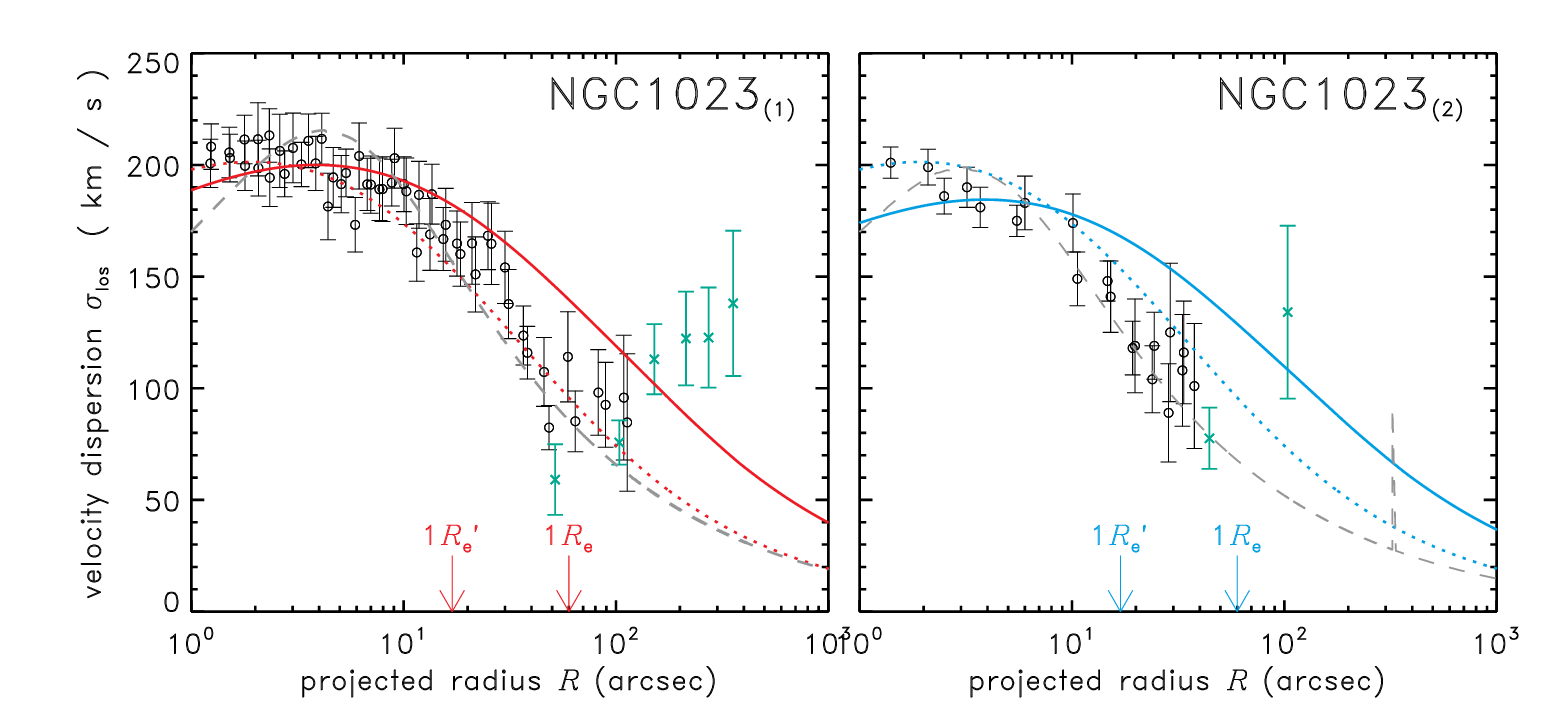}
\\
 \includegraphics[width=7cm]{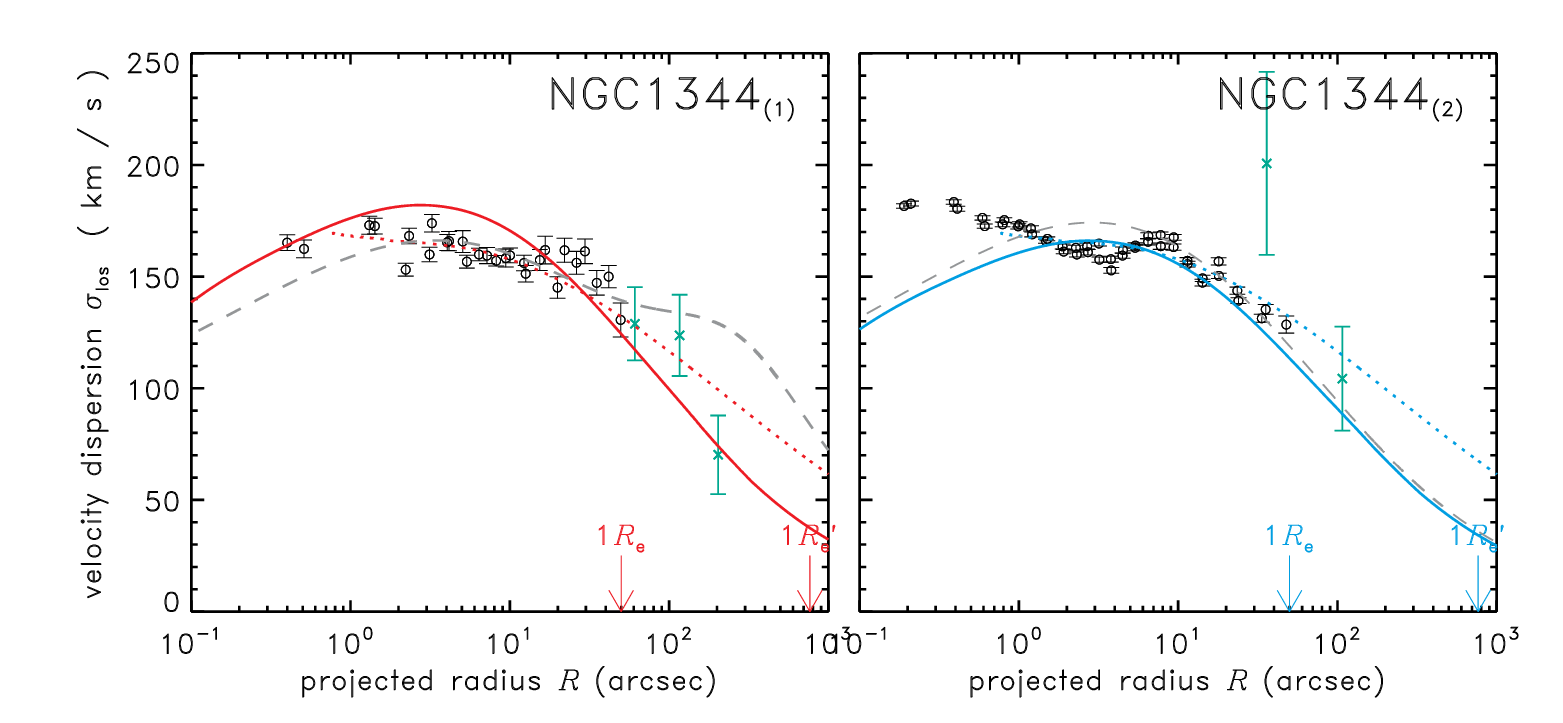}
&\includegraphics[width=7cm]{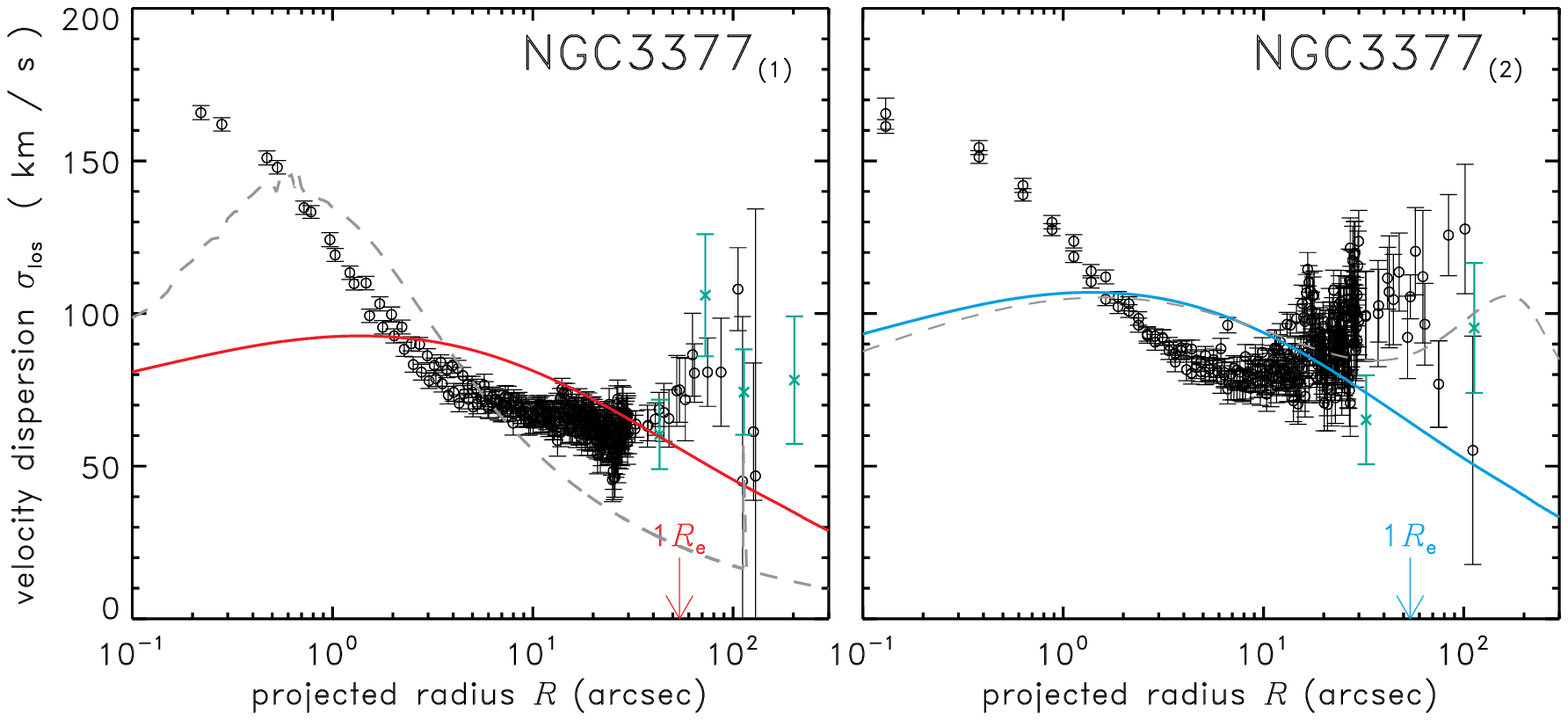}
\\
 \includegraphics[width=7cm]{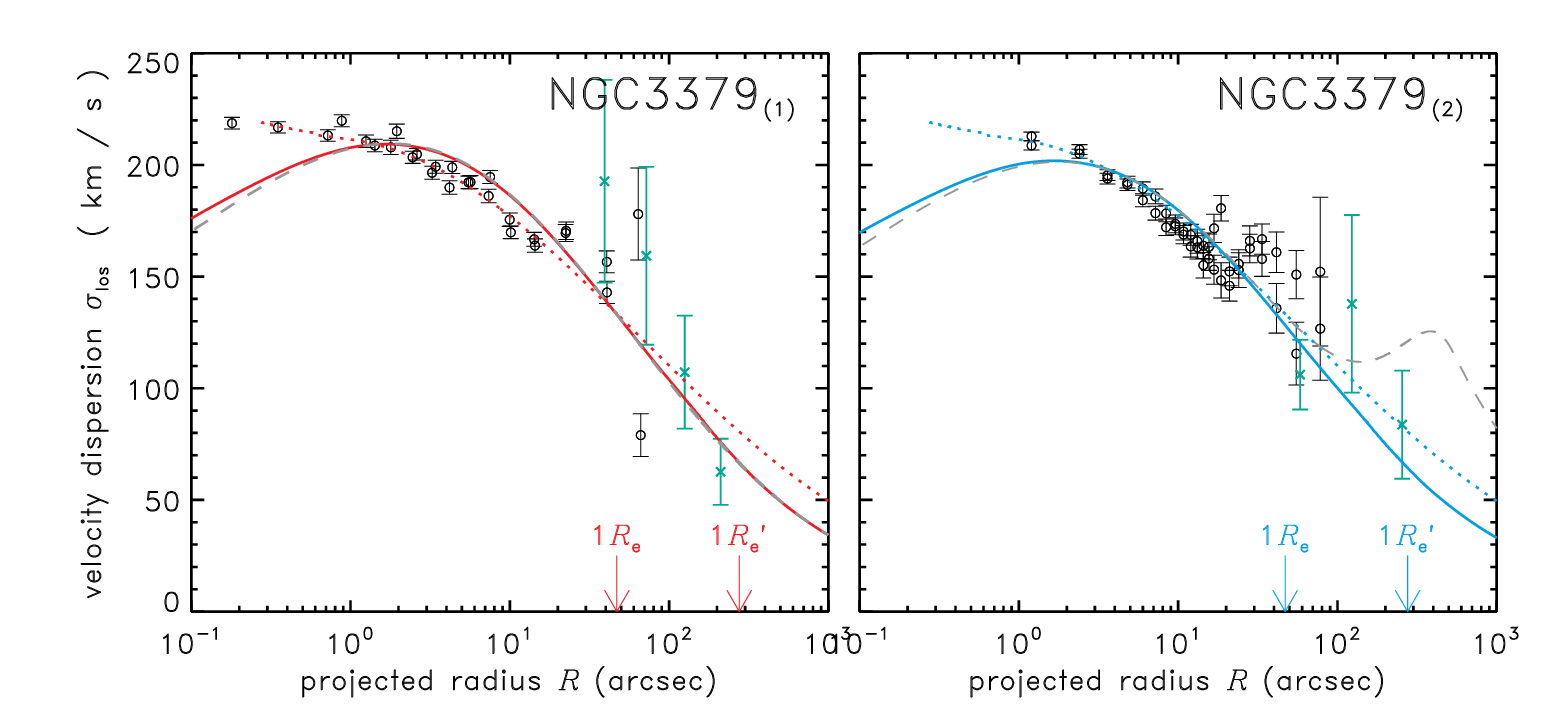}
&\includegraphics[width=7cm]{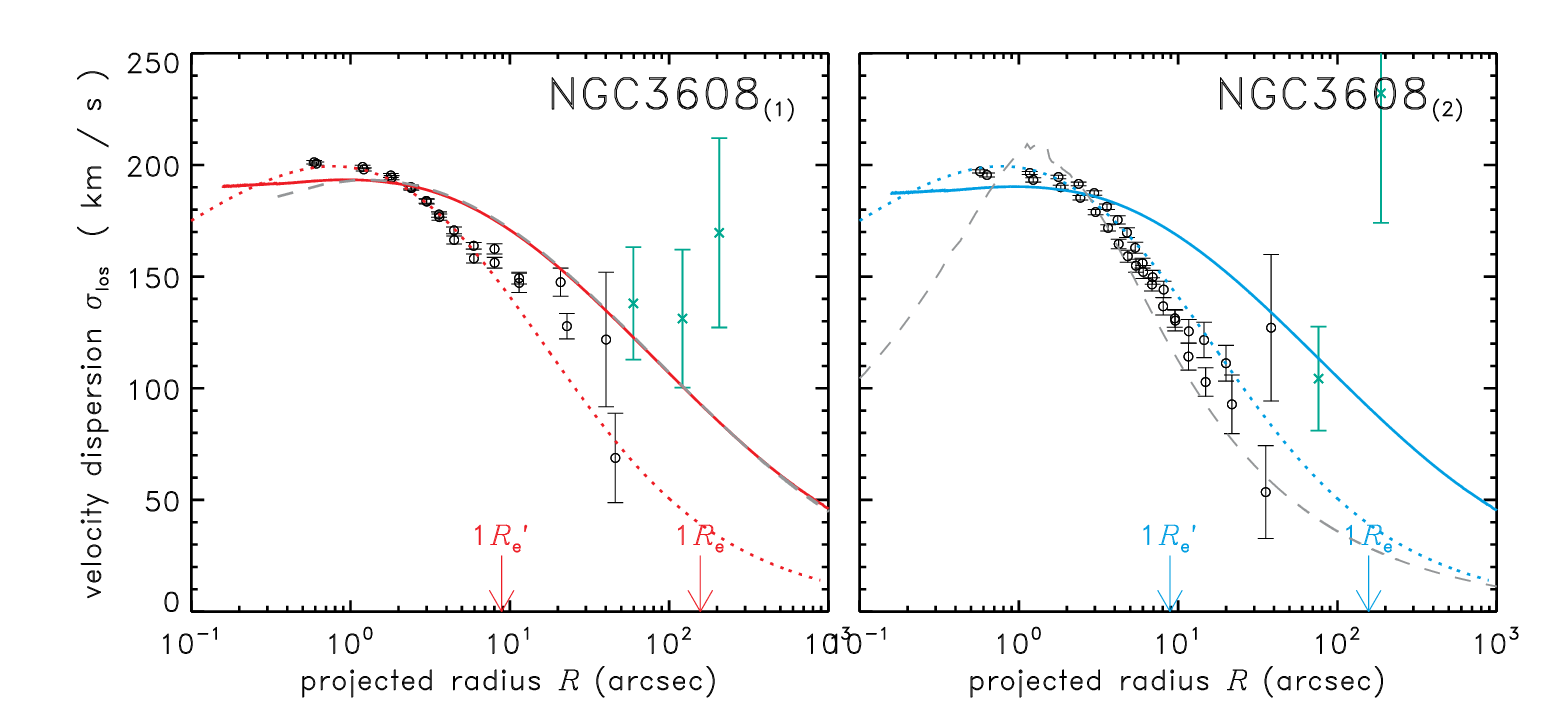}
\\
 \includegraphics[width=7cm]{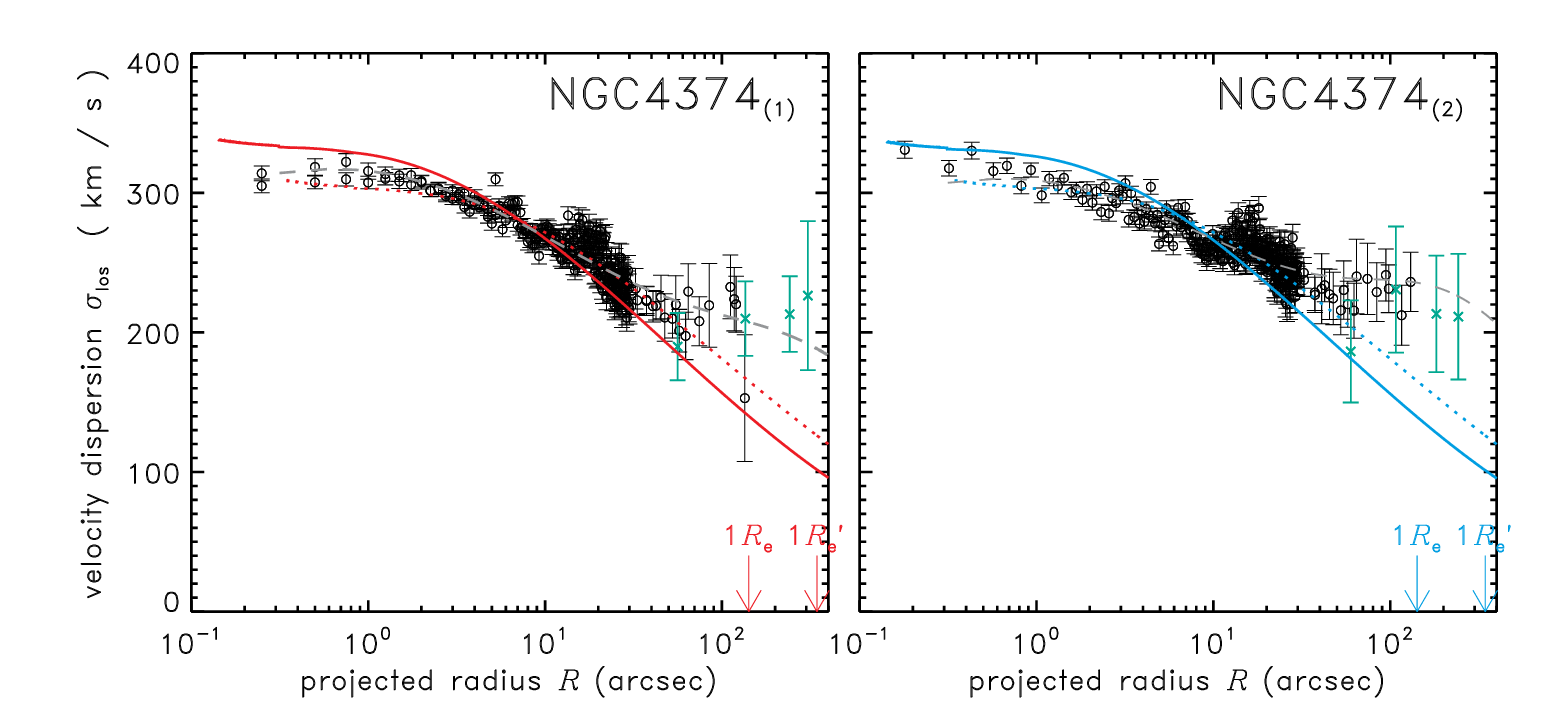}
&\includegraphics[width=7cm]{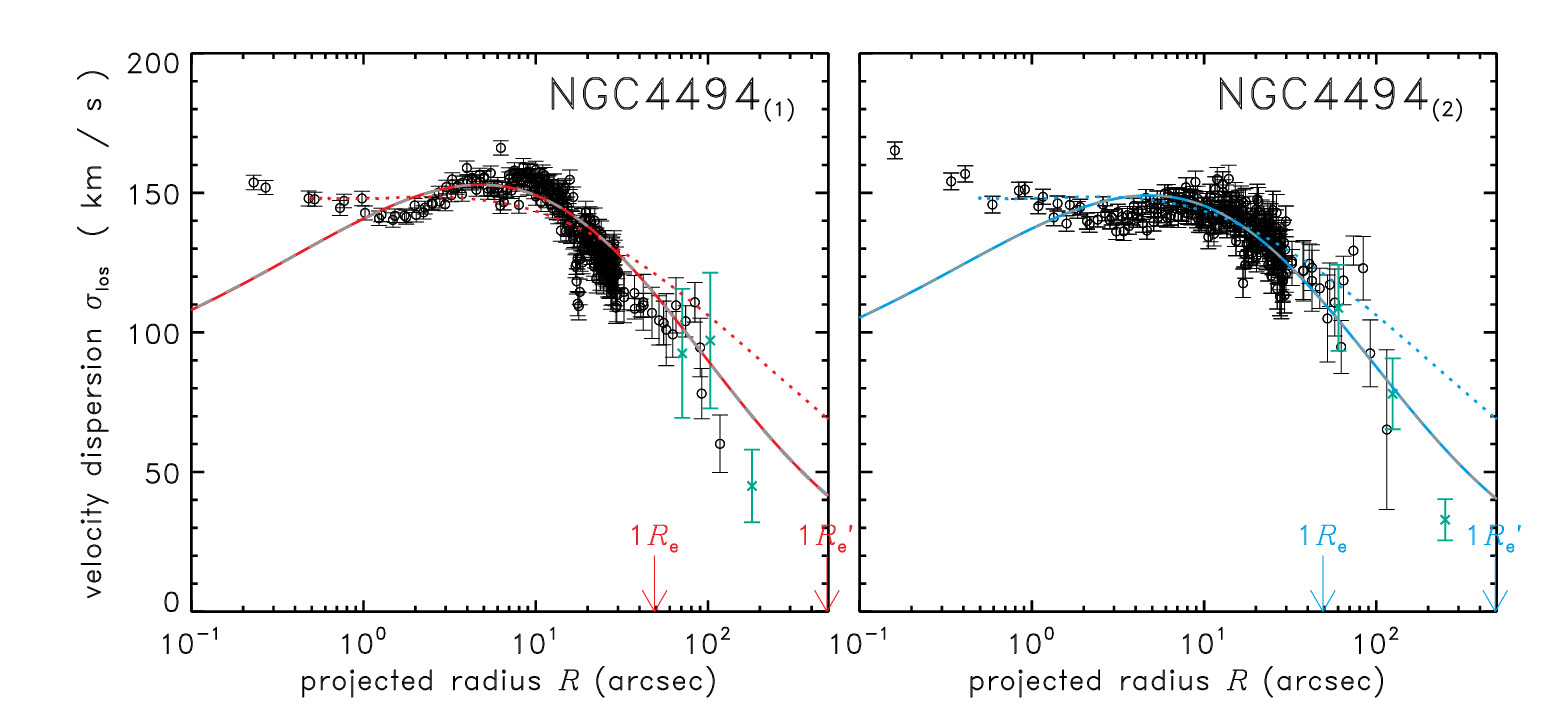}
\\
 \includegraphics[width=7cm]{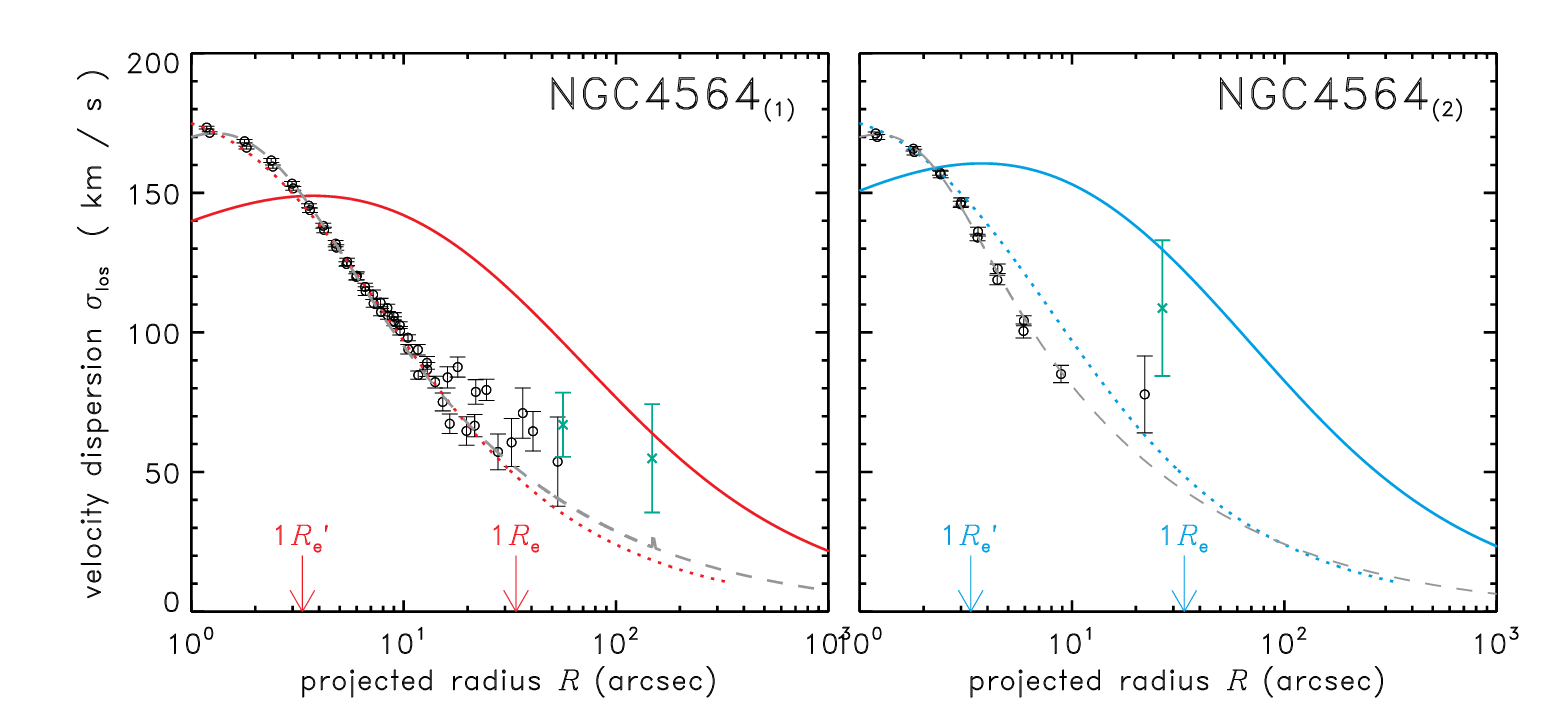}
&\includegraphics[width=7cm]{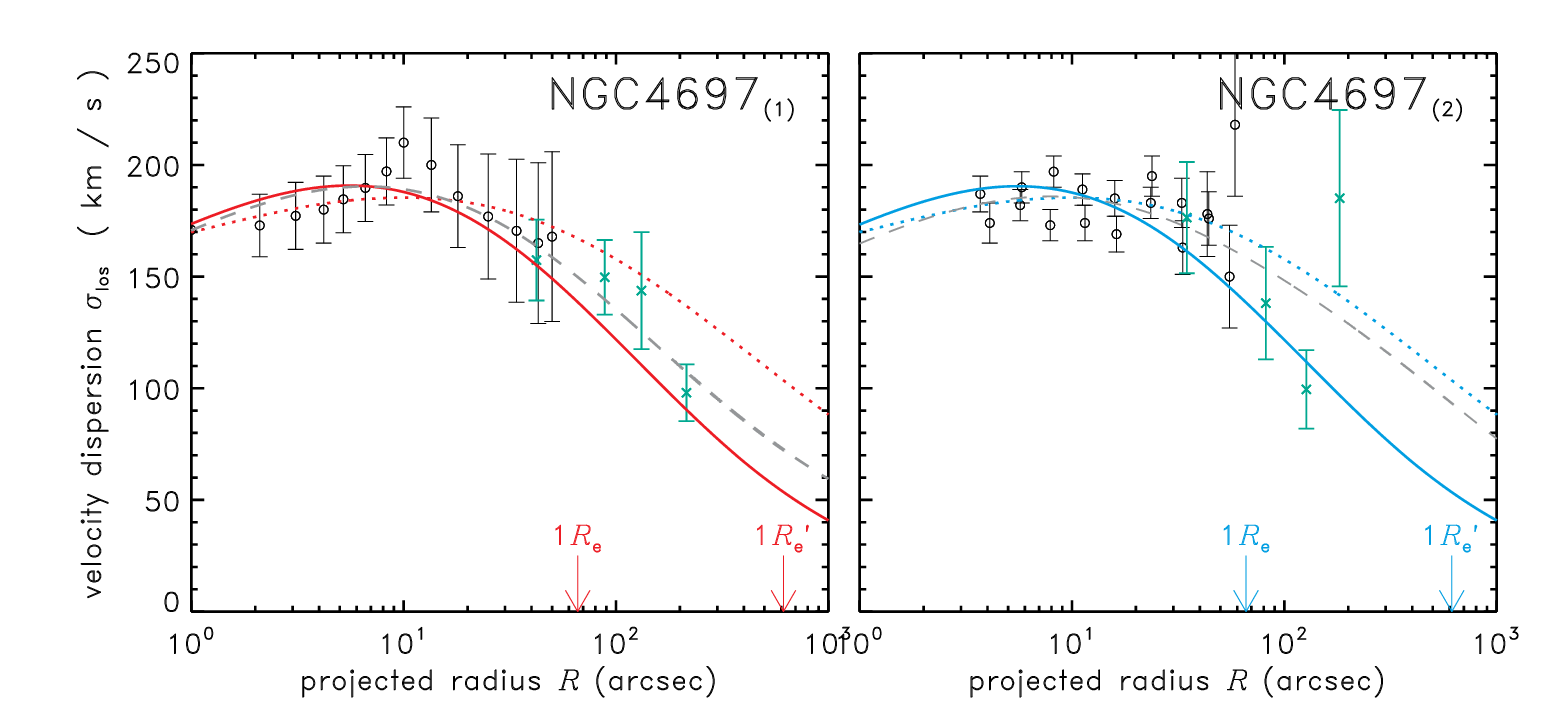}
\\
 \includegraphics[width=7cm]{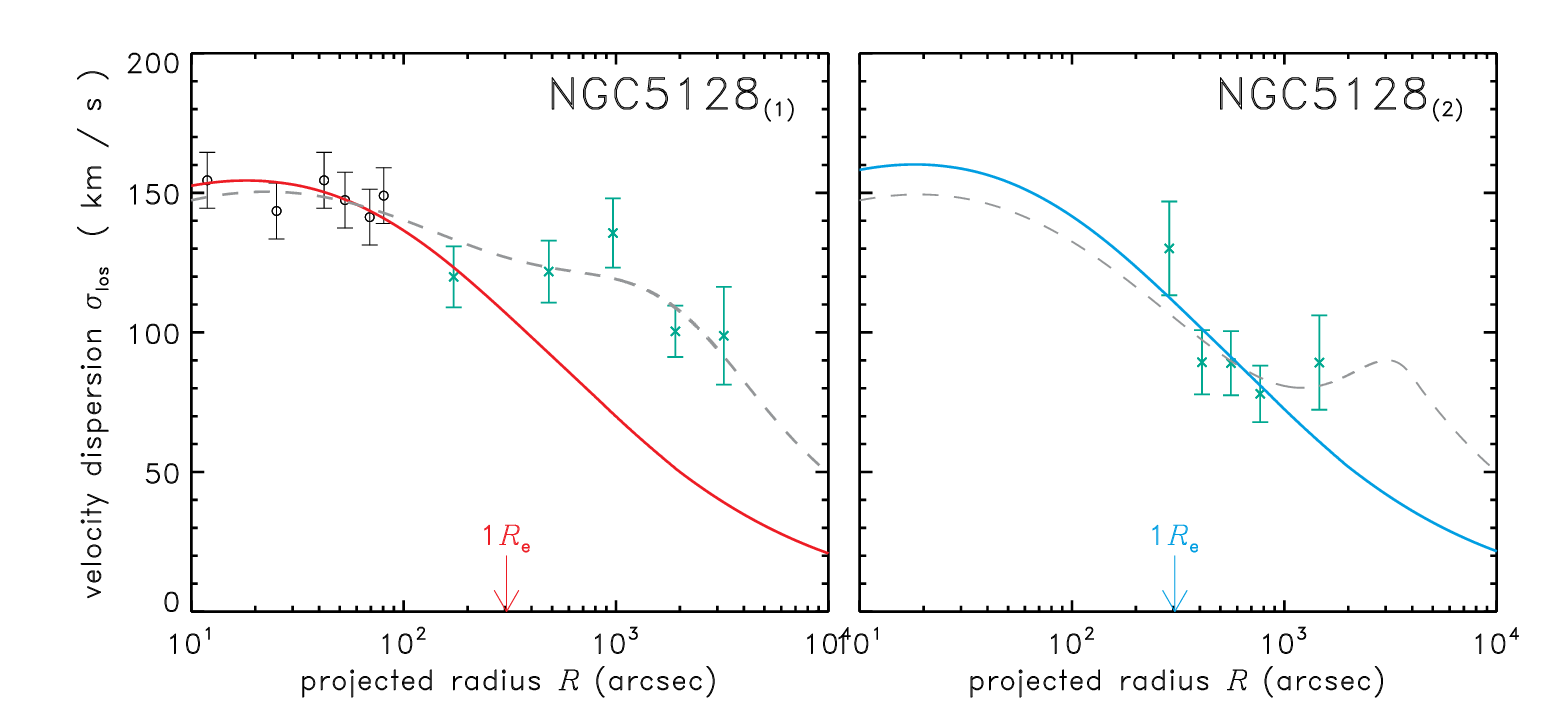}
&\includegraphics[width=7cm]{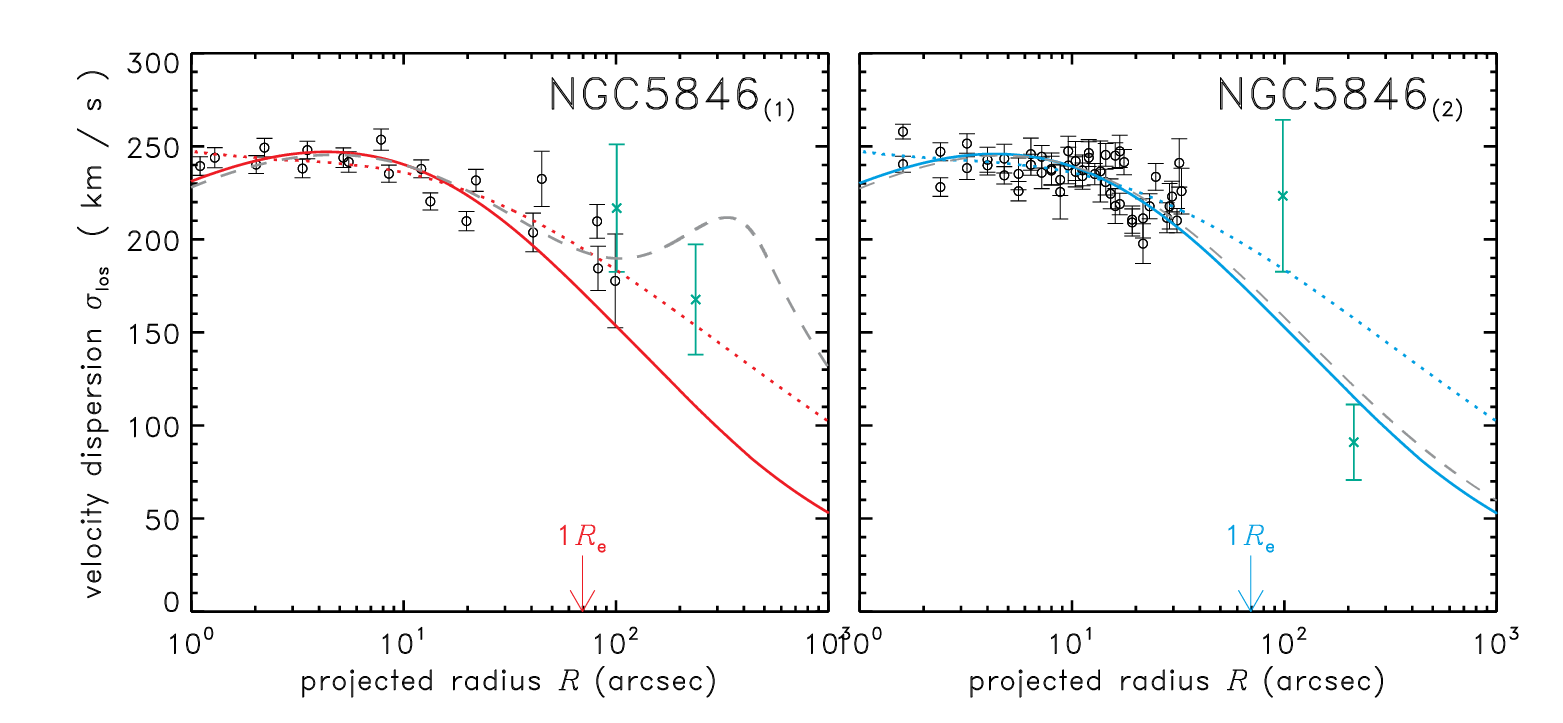}
\\
\end{tabular}
\end{center} 
\caption{Projected velocity dispersion profiles of the 
isotropic models that best fit the \citet{coccato2009} data (on two
principal axes, red and blue).  Stellar data ($\Circle$)
determine these ``stars only'' fits;
the PN ($\times$) data are shown for reference. 
Solid lines are ``stars only'' models based on the standard photometric
S\'ersic parameters; dotted lines reflect parameters adapted to fit
the kinematics better. The dashed lines are the best fits of the
model including a polytropic dark matter halo. On the horizontal
axis, the observed effective radius and the one obtained from the
kinematic fit are shown as $R_{\rm e}$ and $R^\prime_{\rm e}$, respectively 
(for some galaxies the ``kinematic'' $R_{\rm e}$ extends beyond the
range of the figure, and for three cases (NGC~1407; 3377 and 5128) the
kinematic method does not converge; see text for details).}
\label{fig.nodark.fits}
\end{figure*}

Previous modeling of PN kinematics have suggested a poverty of dark
matter in the halos of some isolated elliptical galaxies
\citep{mendez2001,romanowsky2003} or that the halos are less concentrated than
popular collisionless dark matter cosmogonies predict
\citep[e.g.][]{douglas2007,napolitano2009,coccato2009}.  To provide
baselines for comparison, we first fit the Coccato et~al. data with a
purely S\'ersic model (using the published, photometrically obtained
parameters) and no dark matter.  Except for NGC~4697, the goodness of
fit scores -- given by the reduced $\chi_r^2\equiv\chi^2/\nu$ -- are
well above one, and therefore bad fits (see columns 4 and 5 of
Table~\ref{table.stars.optima} and the solid lines in
figure~\ref{fig.nodark.fits}).  The model profiles of the projected velocity
dispersion are smoothly declining outside a half-light radius, whereas
the observational profiles may be flat, or they may dip and rise with
radius (Figure~\ref{fig.nodark.fits}).  Some of these complexities
and departures from ideal Sersic mass profiles may be responsible for
the formally poor fits.

\subsubsection{Improved stellar fits}

Using the photometric S\'ersic parameters tabulated by
\citet{coccato2009}, some galaxies give much worse values for the
reduced $\chi^2$ than others.  In order to improve these cases, we try
fitting isotropic S\'ersic models without dark matter, and adapt the
parameters ($n,R_{\rm e})$ for a better kinematic fit,
($n',R_{\rm e}')$.
In this search, we omit the planetary nebulae, relying on
absorption line kinematics only.

Table~\ref{table.stars.optima} gives the alternative fits for axes 1
and 2 of each galaxy, before and after kinematic fitting.  The left
columns are based on \citet{coccato2009} photometric parameters; the
right columns (primed quantities) are our kinematic fits. We give the
quality of fit $(\chi^2_r)_{\{1,2\}}$ and the total (stellar) mass
$m_{\bigstar_{\{1,2\}}}$, where the subindex refers to the axis
considered in the analysis. These fits are shown in
figure~\ref{fig.nodark.fits} as dotted lines. In some cases we use
subindex 0 to refer to data that were obtained within circular
annuli. We use an ``amoeba'' algorithm to find better fitting
parameters, $(n^\prime,R_{\rm e}^\prime)$, with considerably
improved $(\chi^2_r)^\prime$ on each axis.  For NGC~3377, 1407 and 5128 the
amoeba runs away forever along a degenerate $\chi^2$ valley, heading
towards $R_{\rm e}\rightarrow\infty$.

In the majority of attempts at $(n',R_{\rm e}')$ kinematic fits, we
find $R_{\rm e}$ drifting to large values beyond the data, 
resulting in unphysically large total stellar masses. NGC~4564 and
3608 have smaller $R_{\rm e}'$ than the photometric values.  NGC~1023
is the only galaxy where the photometric and kinematic fitted
$R_{\rm e}$ values are moderate and comparable.  Disagreements of photometric
and kinematic S\'ersic parameters might indicate spatially varying
stellar mass/light ratio, or the existence of dark matter.
Alternatively, the disagreements may be influenced by degeneracies
between the parameters, and the incomplete, finite radial range of the
data.

\subsection{Fitted halos with isotropic orbits}

For the \citet{coccato2009} galaxies, data are available along two
principal axes.  We fit these independently, using a stellar mass
profile with S\'ersic model parameters adopted from the published
stellar photometry (i.e. $n$ and $R_{\rm e}$
from table~\ref{table.stars.optima}). 
The central properties are tuned so that the dark
halo and stars are globally in a chosen ratio.  We explore cases with
fixed $F_{\rm d}$, and vary the halo radius (via $s_{\rm d}$) to
achieve the best fit.  Comparing the fits for the two axes gives some
indication of the quality of the spherical approximation.  (It is an
imprecise comparison however, since the minor axis data tend to span a
shorter radial range, and are less restrictive.)  For six galaxies
there are formally satisfactory fits, with $\chi^2_r\la1$ (i.e. in
NGC~1407, 4374, 4564, 4697, 5128 and 5846).  For the rest, the minimal
$\chi^2_r$ never drops below a few (formally unfavourable).  This
may be partly due to ideosyncratic stellar kinematic substructures or
noisiness.

\begin{figure*}
\begin{center} 
\begin{tabular}{ccc}
	 \includegraphics[width=5cm]{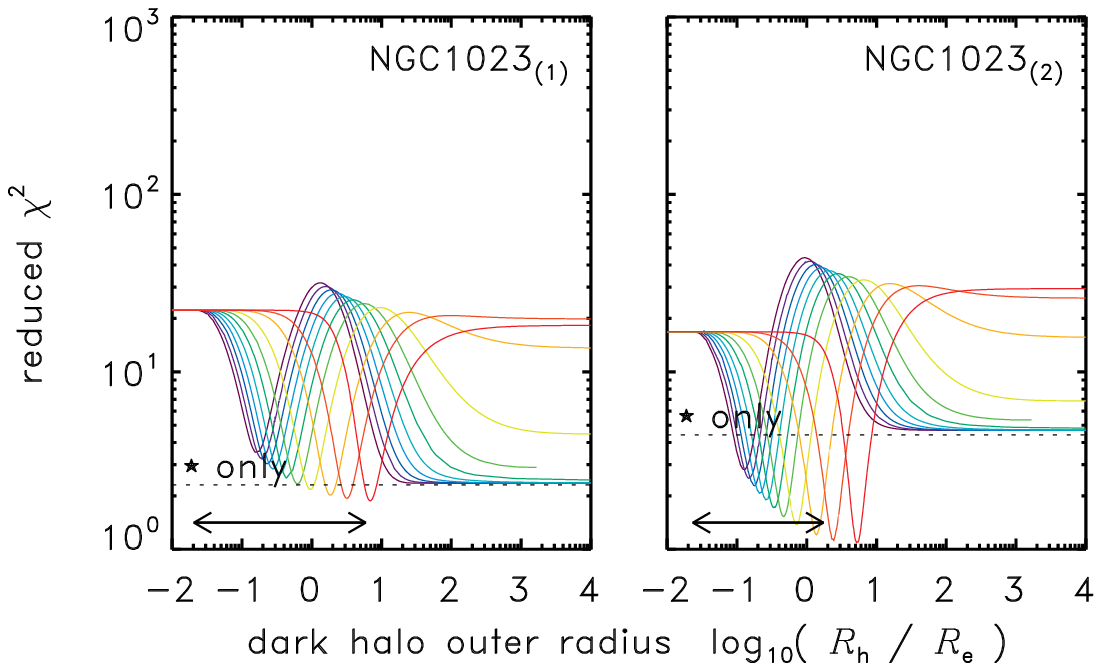}
	&\includegraphics[width=5cm]{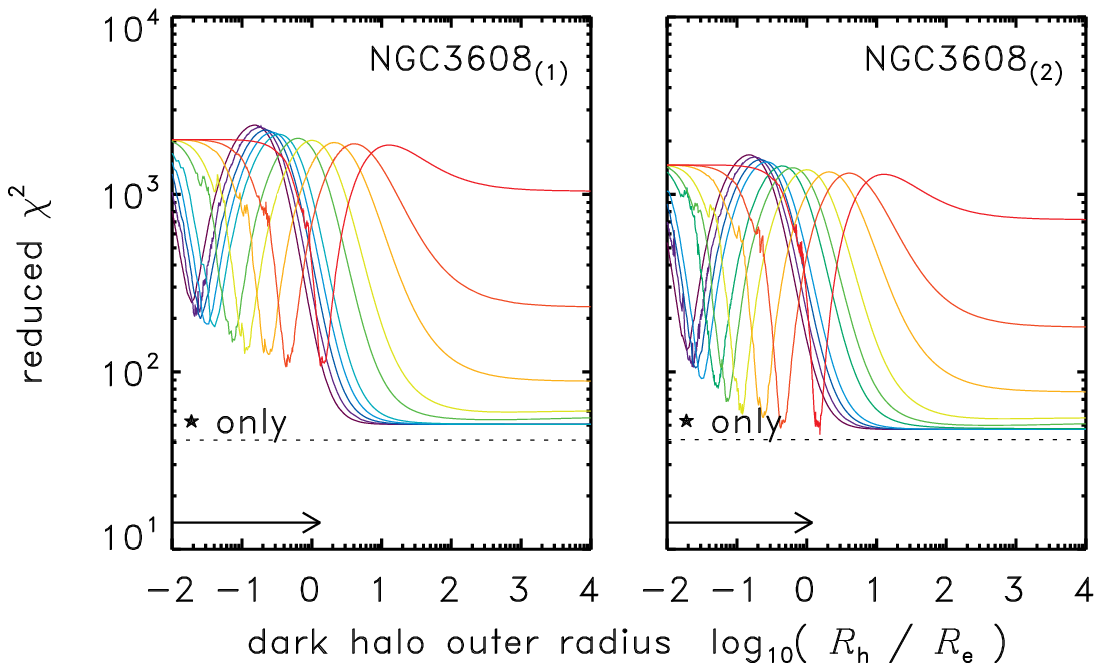}
	&\includegraphics[width=5cm]{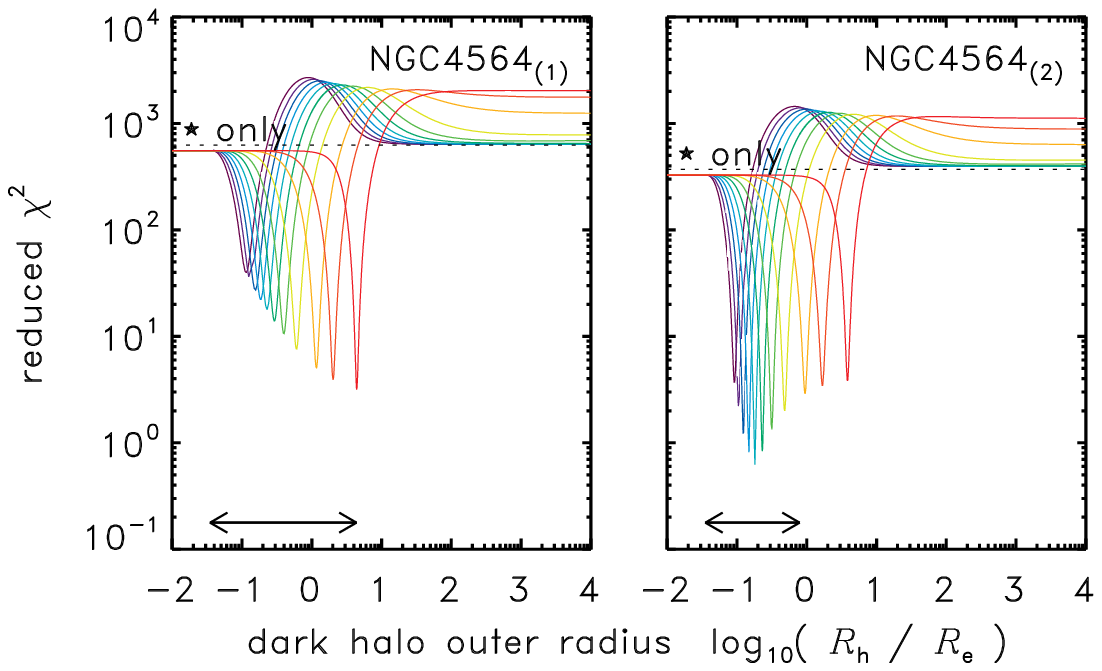}
	\\
	\\
	 \includegraphics[width=5cm]{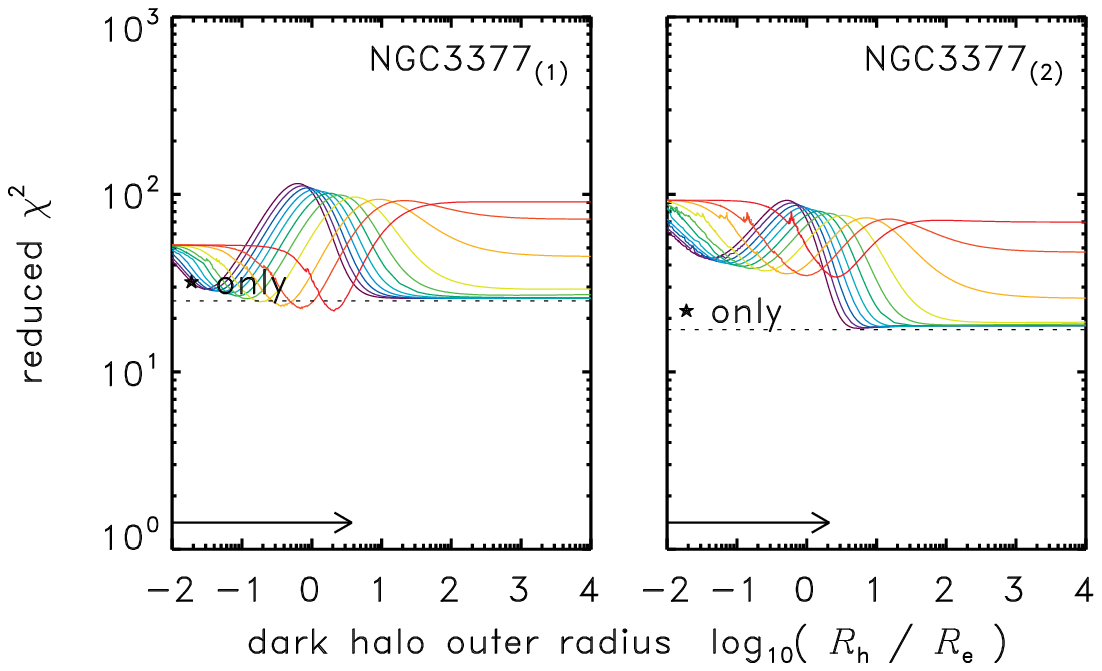}
	&\includegraphics[width=5cm]{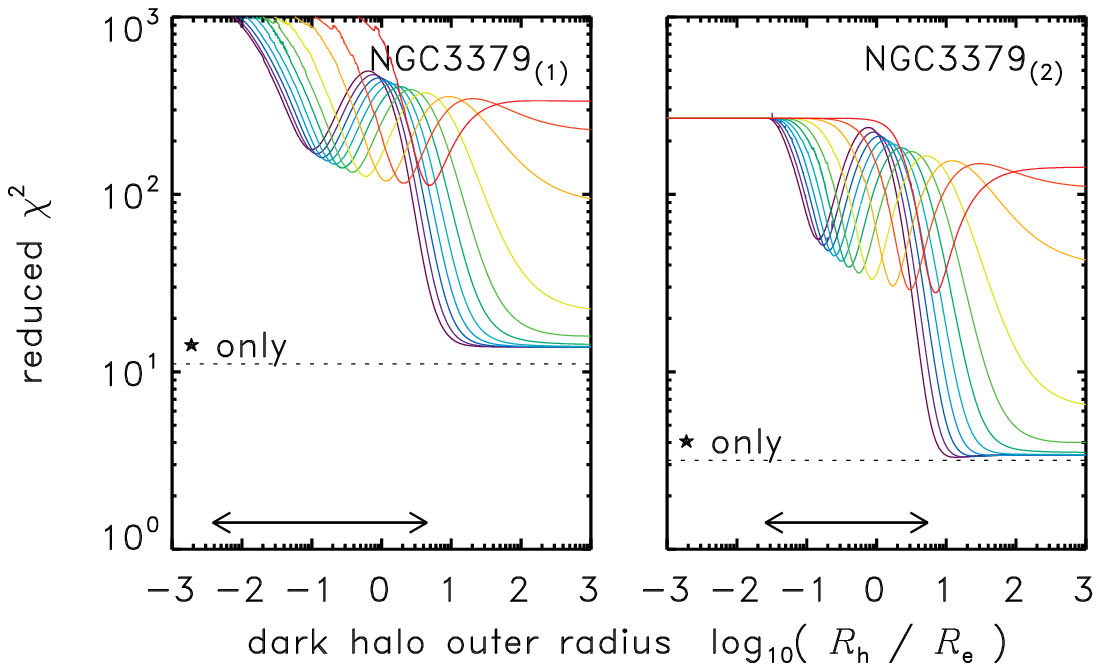}
	&\includegraphics[width=5cm]{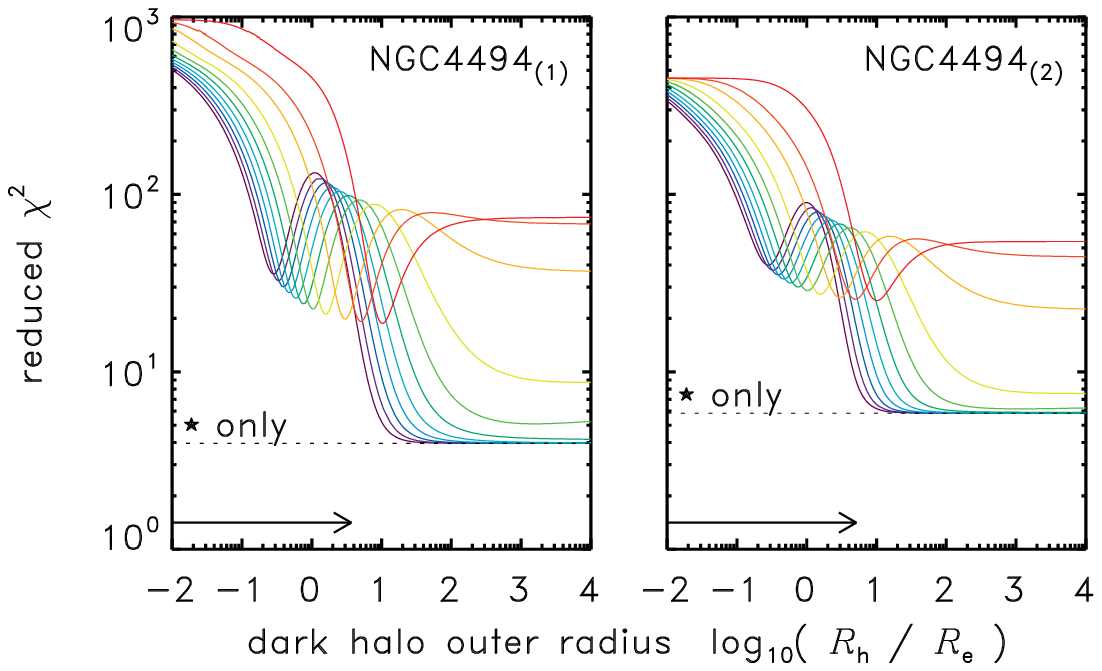}
	\\
	\\
	 \includegraphics[width=5cm]{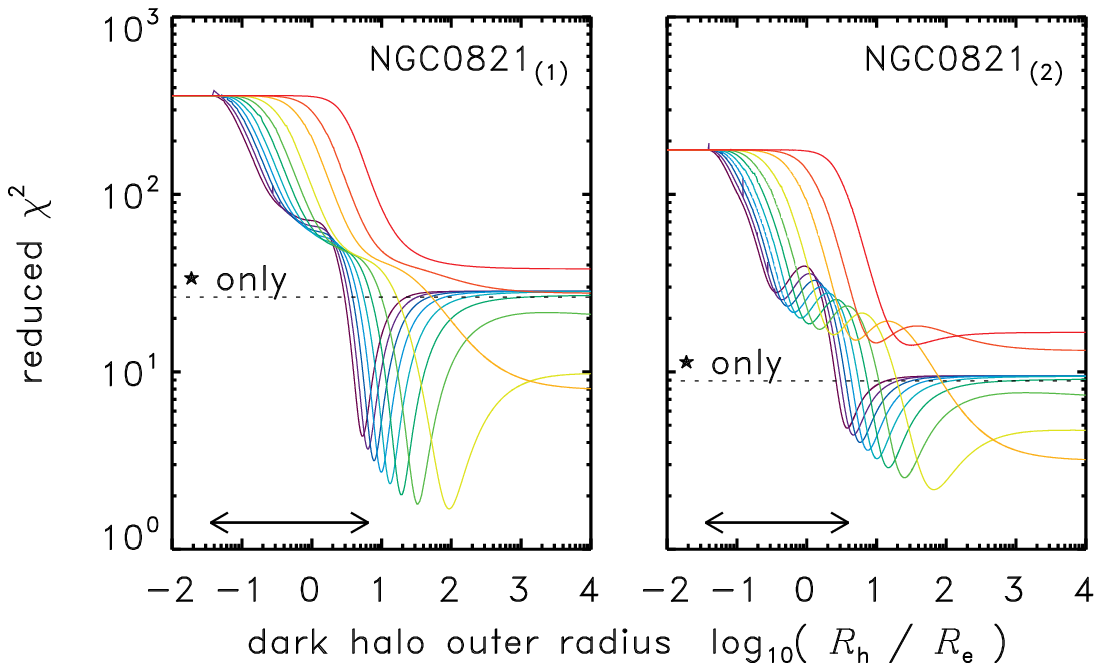}
	&\includegraphics[width=5cm]{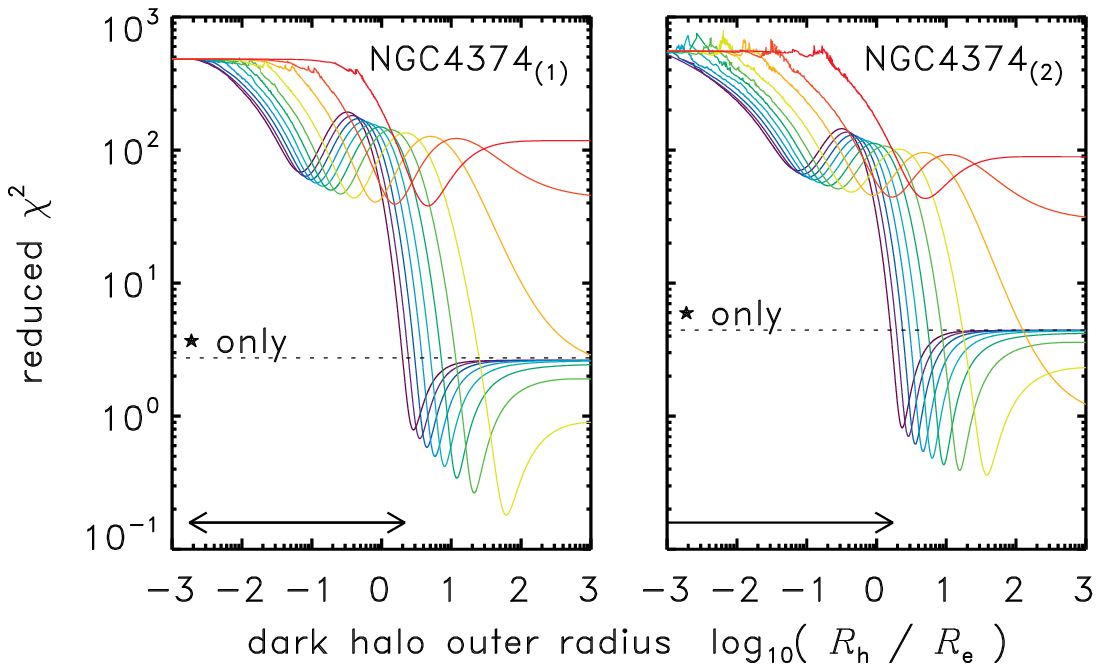}
	&\includegraphics[width=5cm]{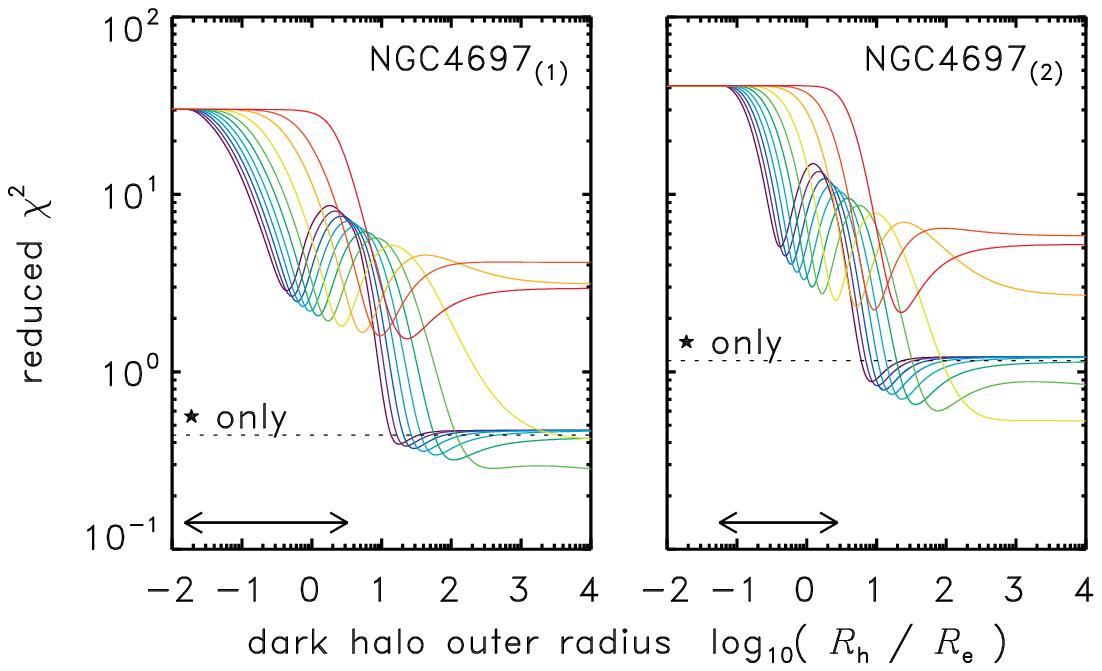}
	\\
	\\
	 \includegraphics[width=5cm]{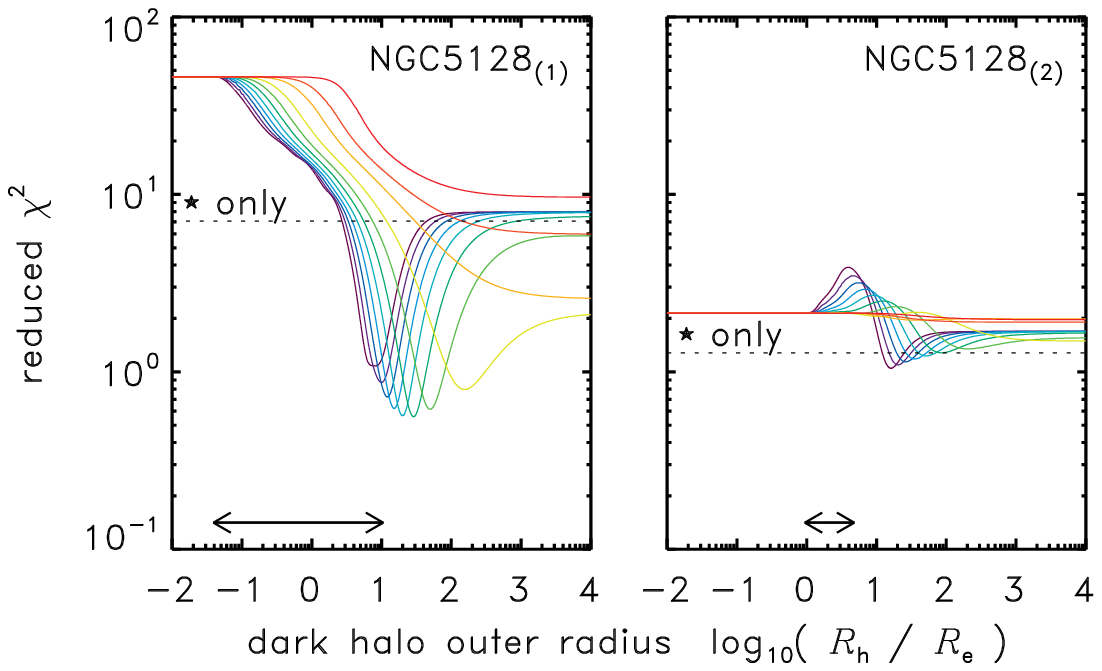}
	&\includegraphics[width=5cm]{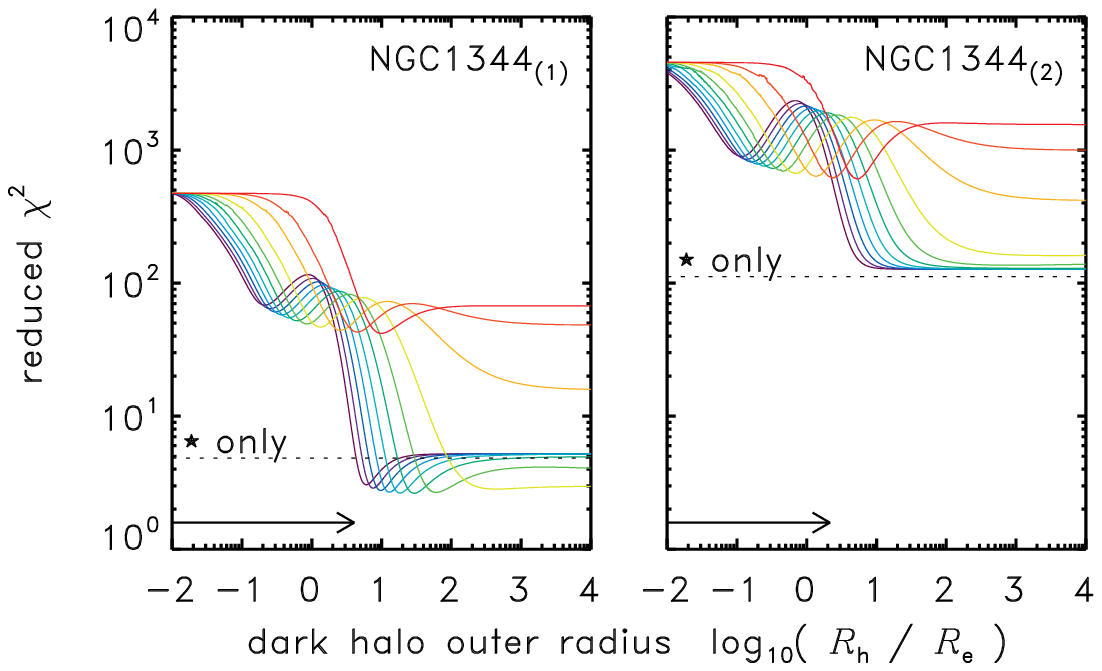}
	&\includegraphics[width=5cm]{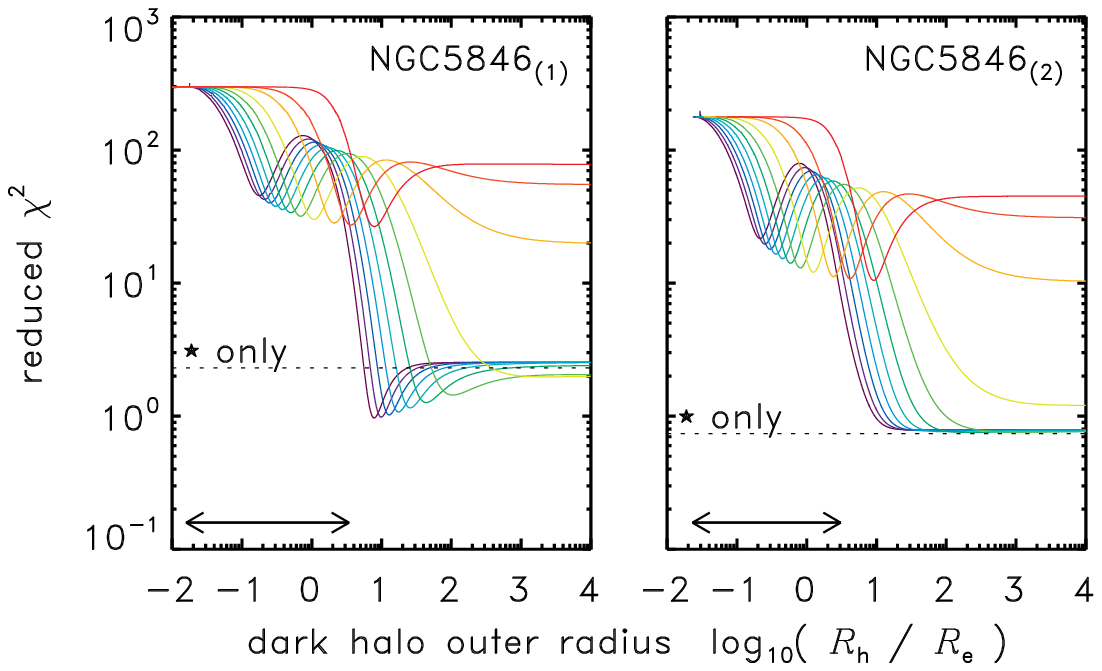}
\end{tabular}
\end{center} 
\caption{
Quality of isotropic fits to the
Coccato et~al. kinematic data for two principal axes,
as a function of halo outer radius $R_{\rm h}$.
Colours represent cases of
different degrees of freedom,
$F_{\rm d}=1,2,3,4,5,6,7,8,9,9.5,9.9$ 
as indicated in the legend in figure~\ref{fig.isotropic2}.
Dotted lines indicate the quality of fit for the S\'ersic only model.
Arrows show the radial coverage of kinematic data.
Stellar data have been softened
to a best-fit cubic.
Unusually, some of the $\chi^2$ minima of NGC~1023, 3608 and 4564
imply very compact halos
($R_{\rm h}\la R_{\rm e}$)
fitting better than the ``stars only'' level.
More plausible fits (lower two rows) prefer a halo radius
outside the observed stars and PN.
}
\label{fig.isotropic}
\end{figure*}

\begin{figure*}
\begin{center} 
\begin{tabular}{ccc}
	 \includegraphics[width=5.9cm]{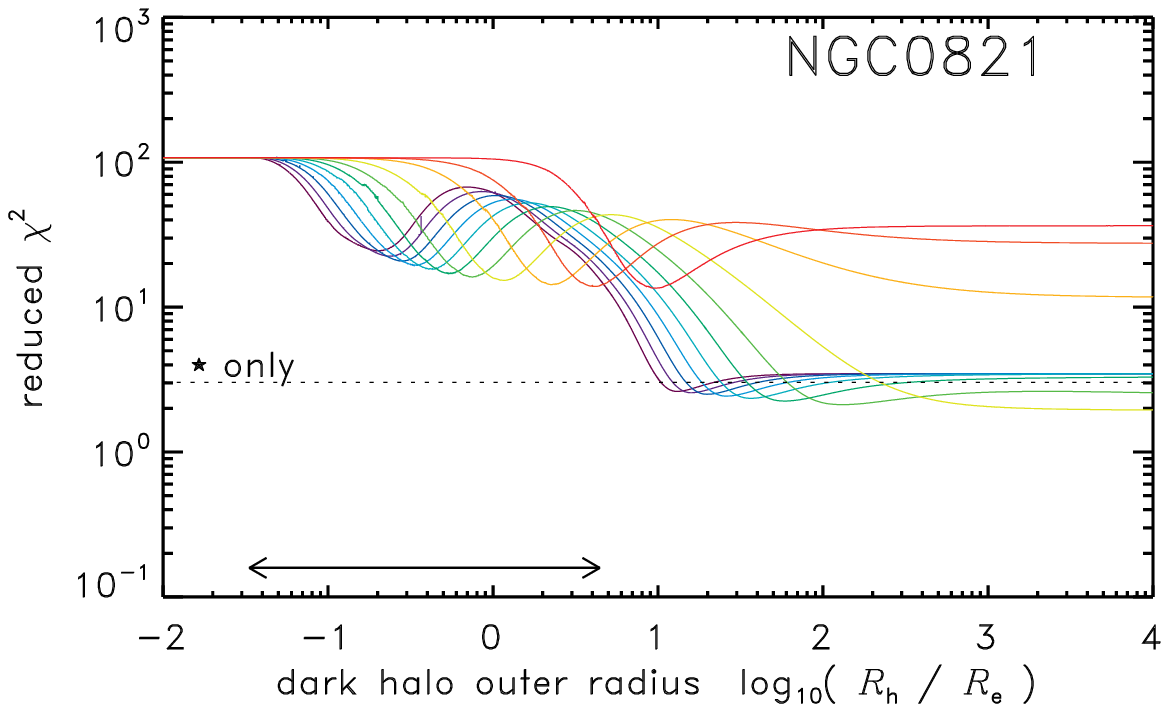}
	&\includegraphics[width=5.9cm]{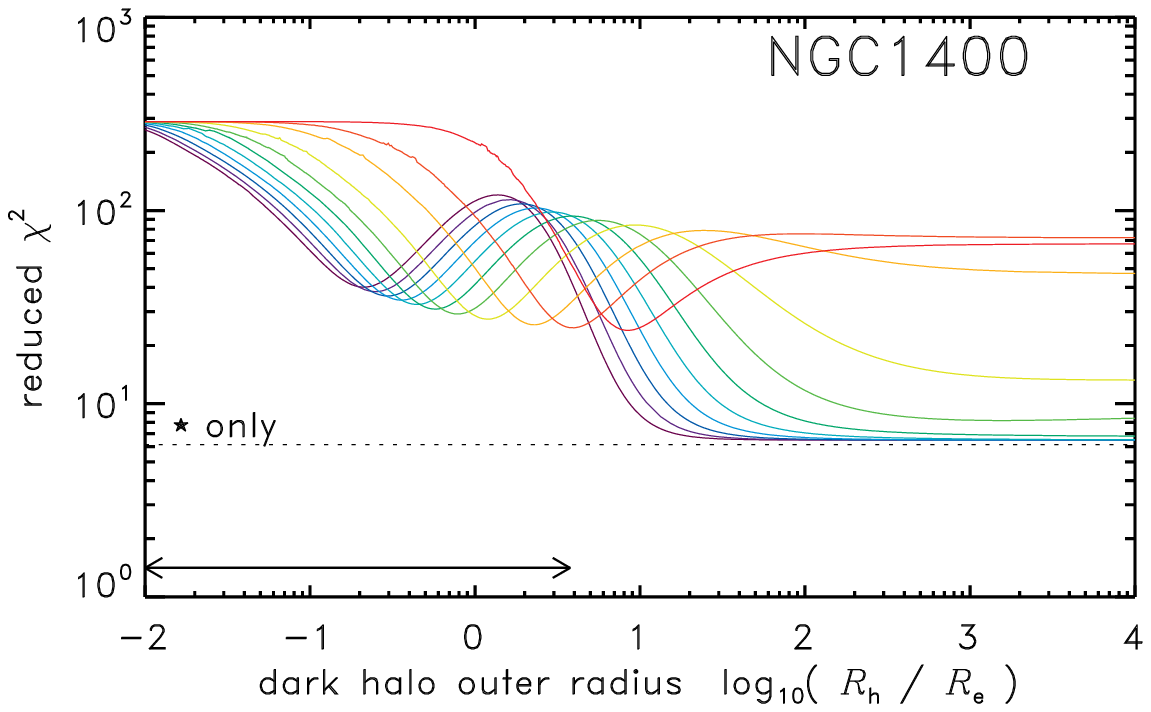}
	\\
	 \includegraphics[width=5.9cm]{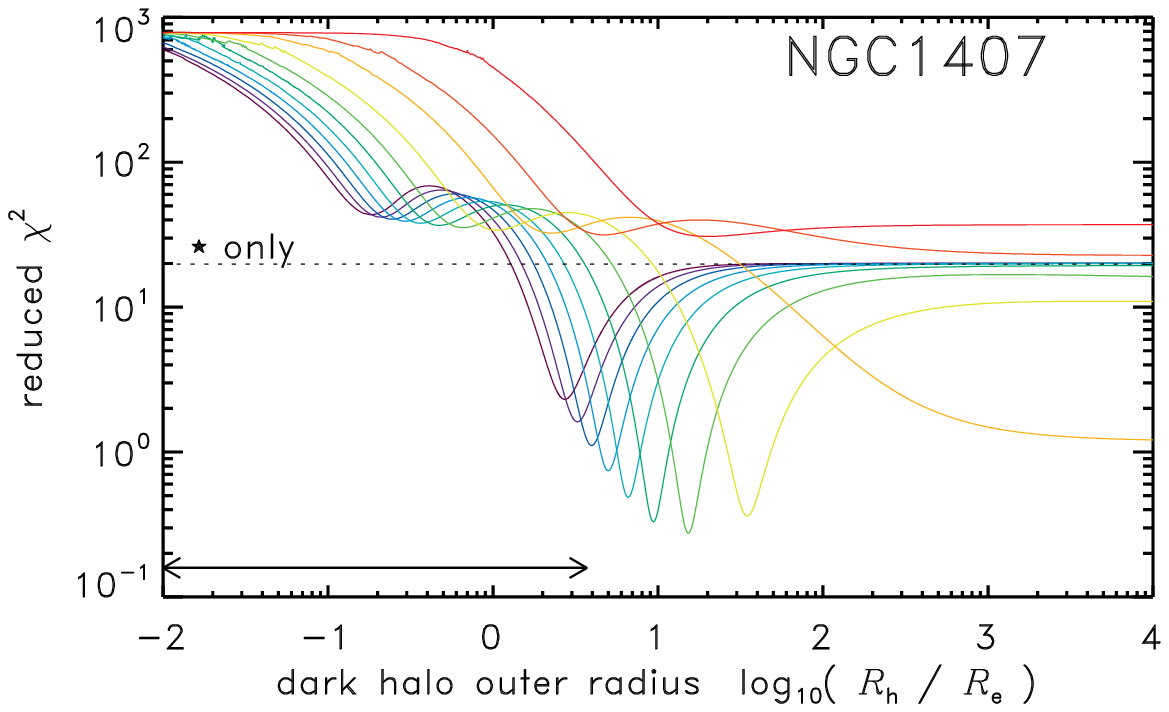}
	&\includegraphics[width=5.9cm]{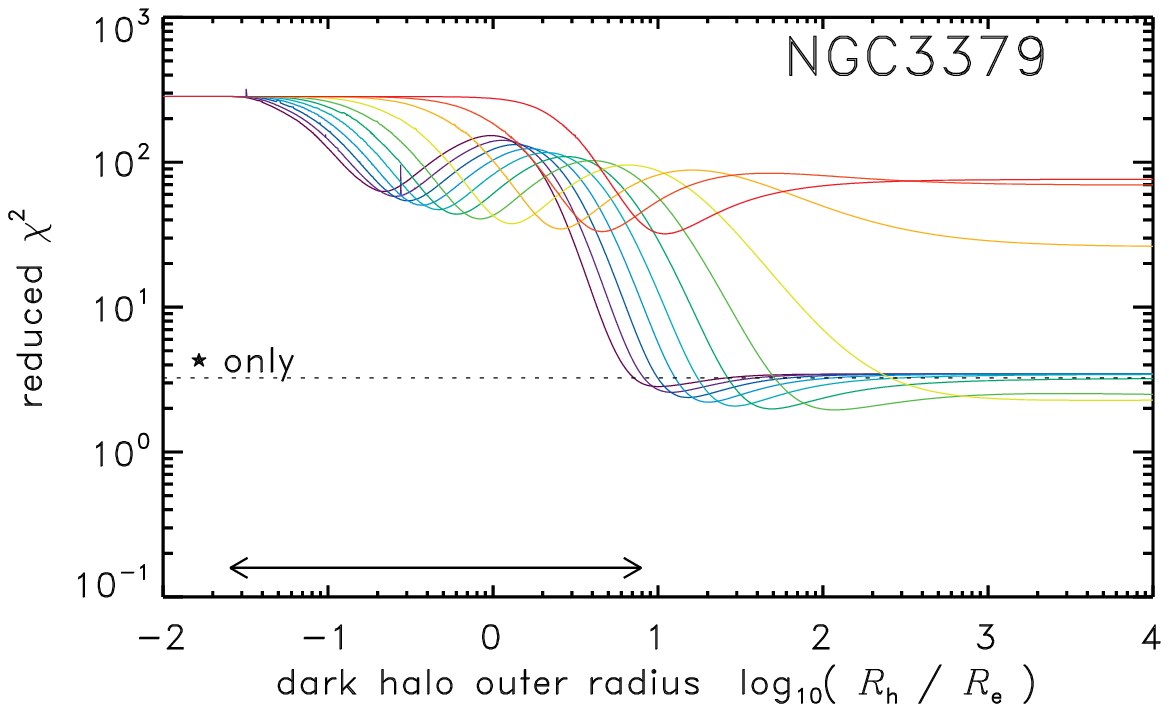}
	&\includegraphics[width=4.5cm]{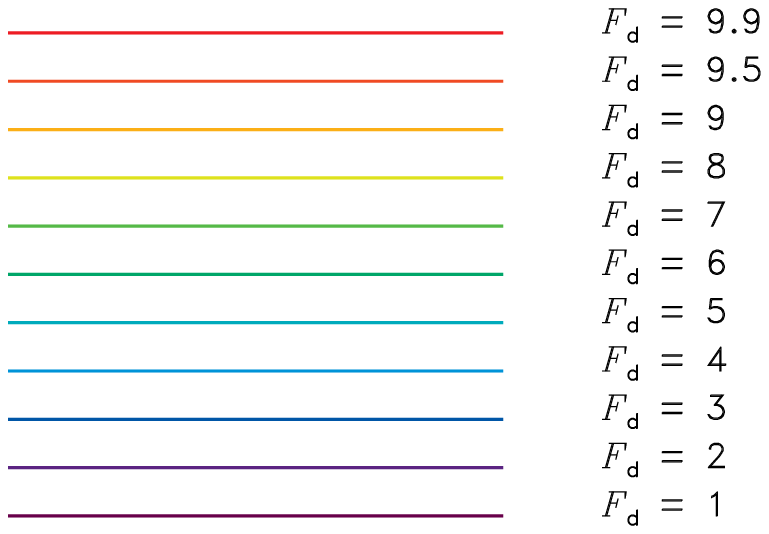}
\end{tabular}
\end{center} 
\caption{
Same as figure~\ref{fig.isotropic} for data without axial splitting
\citep{romanowsky2003,douglas2007,proctor2009}.
}
\label{fig.isotropic2}
\end{figure*}

To address this possibility, we repeat the kinematic fitting process
with the stellar $\sigma_{\rm los}(r)$ profile smoothed to a best-fit cubic
function in $\ln(r)$.  PN data are left untouched.  Smoothing the
stellar kinematics does not shift the minima significantly, but lowers
the attainable $\chi^2_r$.  Thus the minima are sharper and deeper,
but the landscape is unchanged at medium and high $\chi^2$ levels.
Results for the smoothed fitting are shown in
Figures~\ref{fig.isotropic} and \ref{fig.isotropic2}. The colours from
violet to red represent dark degrees of freedom $F_{\rm d} =
\{1,2,3,...7,8,9,9.5,9.9\}$.  The PN velocities were assumed to be
isotropic.  We assume a universal cosmic fraction of dark and baryonic
mass (i.e. stars for our galaxies) according to the WMAP5 cosmology
\citep{wmap5}. In figure~\ref{fig.isotropic2} we show four galaxies for
which we have data within circular apertures \citep{romanowsky2003,douglas2007,proctor2009}.

The usual $\chi^2$ curve of a galaxy has two minima: a shallow dip
around $R_{\rm h}\sim0.1R_{\rm e}$ to 1$R_{\rm e}$, and a deeper dip at radii
outside the observations.  The physical interpretation of the inner
dip implies a dark halo more compact than the stellar distribution.
This would be surprising in the conventional cosmogonies in which
radiative cooling made the star-forming gas more concentrated than the
dark component.  Numerically, at inner-dip configurations the ``halo''
is taking over the role of the stars influencing the inner
$\sigma_{\rm los}$ profile.  The outer minima seem more physically
plausible.  Sometimes the outer dip is effectively a lower limit on
$R_{\rm h}$ (e.g. for NGC~4494) but in other cases the fitting prefers a
finite range (e.g. less than 100$R_{\rm e}$ for NGC~821).  If the data
are clear and widespread enough, the radial variation of the slope of
$\rho_{\rm d}$ in the observed range is sufficient to distinguish
$F_{\rm d}$ and extrapolate the truncation radius.
For $F_{\rm d}<9$ the $\chi^2(R_{\rm h})$ curves
tend towards the ``stars only'' level
as $R_{\rm h}\rightarrow\infty$.
A large-$R_{\rm h}$ halo (of fixed $\mu_{\rm d}$) is spread thinly
and has negligible effect on observable kinematics.
The convergence is closer for lower $F_{\rm d}$,
as these models resemble the incompressible limit ($F_{\rm d}=0$)
which has uniform density within $r<R_{\rm h}$.
The highly compressible $F_{\rm d}\ga9$ cases behave quite differently,
as discussed in \S\ref{s.features}. 
   
\begin{figure*}
\begin{center} 
\includegraphics[width=5in]{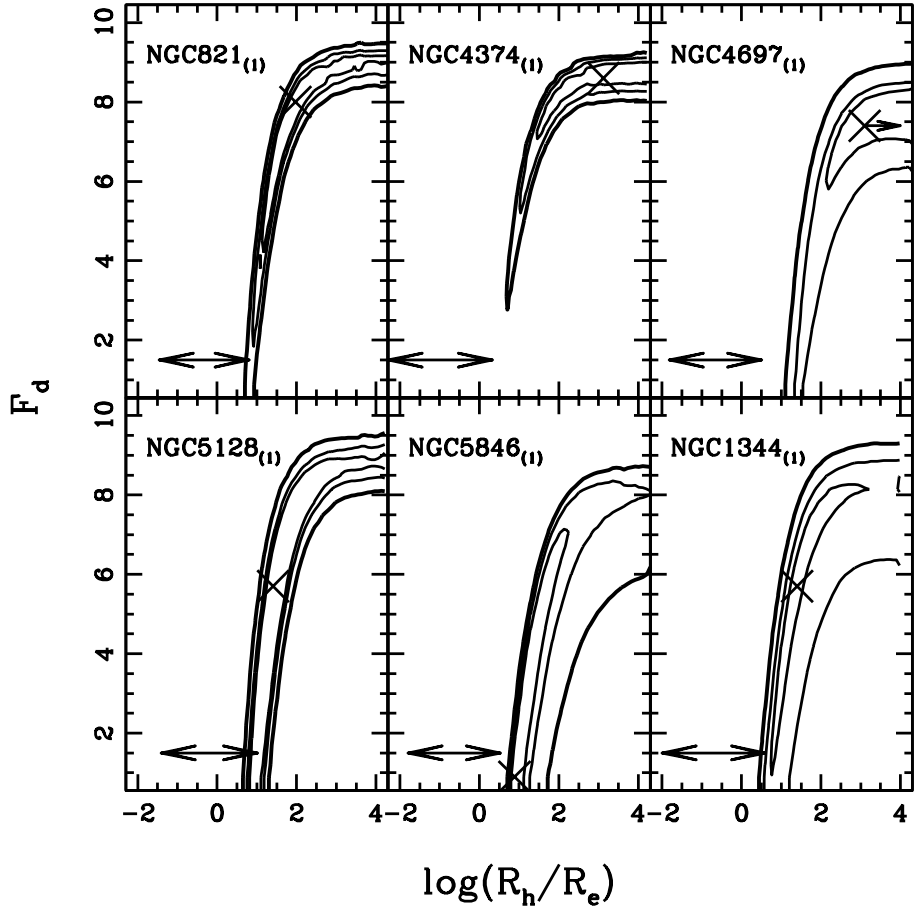}
\end{center} 
\caption{The confidence levels in the parameter space spanned by
truncation radius ($R_{\rm h}$; horizontal) and number of degrees of
freedom of the dark matter equation of state ($F_{\rm d}$; vertical)
are shown for six of the galaxies from our sample. They correspond to
the 90, 95 and 99\% (thick like) level, using the likelihood defined
by the fit to the velocity dispersion data. The horizontal arrow
delimits the range over which the observed data are available.  The
crosses in each panel give the best fit. Notice the best fit for the
truncation radius of NGC~4697 falls outside of the range shown (see
table~\ref{table.mol}).  }
\label{fig.chi2}
\end{figure*}

Figure~\ref{fig.chi2} shows the two-dimensional 90, 95 and 99\%
confidence levels for six galaxies in our sample, in the parameter
space spanned by the dark matter degrees of freedom ($F_{\rm d}$) and the
halo truncation radius ($R_{\rm h}$). The crosses mark the best fit, and
the arrow represents the region over which the velocity dispersions are
fit. The above mentioned degeneracy seen in the 1D plots of
figure~\ref{fig.isotropic} is actually a continuous one between $F_{\rm d}$
and $R_{\rm h}$.

For three exceptional galaxies (NGC~1023, 3608, 4564) the inner
minimum is very significant, or even dominant, in terms of $\chi^2$.
Inspection of the $\sigma_{\rm los}$ profiles shows that the
photometrically-derived S\'ersic model fits them especially poorly.
The inner part of the stellar profile is unusually centrally peaked.
NGC~3377 also has a peculiarly shaped $\sigma_{\rm los}$ profile,
which may be responsible for the slight preference for $F_{\rm d}=9.9$
on the first axis, and for the failure to improve on the stellar-only
baseline fit to the second axis.  The modelling of these four misfit
galaxies implies very small stellar mass to light ratios (in the
$B$-band), typically $\Upsilon^B_\bigstar\la0.1\Upsilon^B_\odot$ (see
NGC~$1023_{(1,2)}$, NGC~$3377_{(1)}$ and NGC~$4564_{(1,2)}$ in
Table~\ref{table.mol}).  This is implausible for the aged stellar
populations of early-type galaxies.  We deprecate the discussion of
these particular galaxies.

Except for these odd cases, the ``stars + halo'' fits tend to reject
$F_{\rm d}\gtrsim 9$.  Where the data are clear and smooth enough, the
isotropic modelling generally prefers the dark halo to have $F_{\rm
d}=8$ or fewer effective thermal degrees of freedom.  There is a mild
preference for the top end of the range.  In models with $F_{\rm
d}\ge9$ the central dark density is insensitive to variations of the
outer halo radius ($R_{\rm h}$) or the halo specific entropy ($s_{\rm d}$).
This inflexibility in fitting central conditions may cause the poor
fits for $F_{\rm d}\ge 9$.  The best fits favour halo radii of at
least tens of $R_{\rm e}$.  

We now comment on the favoured isotropic models for specific galaxies.

\subsubsection{NGC~821}
The kinematics of planetary nebulae around NGC~821 show a radially
declining velocity dispersion \citep{romanowsky2003}.  This has
been taken as evidence disfavouring the presence of a dark halo
(unless the PN orbital anisotropy has significant radial variations)
or else the halo is less
concentrated than CDM expectations.

On both axes of the \citet{coccato2009} data, we find that polytropic
halo models fit significantly better than stars alone.  The
improvement is best on the first axis.  The best fits
($\chi^2_r\approx1.7$) are for $F_{\rm d}\approx8$ and an outer radius
$60\la R_{\rm h}/R_{\rm e}\la90$.  These minima improve significantly upon
the ``stars only'' fit, in the case of axis 1 by up to $\Delta\chi^2_r\sim$25.

The upper left panel of Figure~\ref{fig.isotropic2} presents the equivalent
reduced-$\chi^2$ scores obtained by fitting polytropic dark halos to
the \citet{romanowsky2003} data for NGC~821.  These data were annular
in distribution (not divided onto axes) and less detailed than
\citet{coccato2009}.  The fits also prefer $F_{\rm d}\approx 8$, but
the minima are shallower. For $F_{\rm d}=8$ the fit improves gradually
for larger $R_{\rm h}$, but for $F_{\rm d}\le7$ there is an optimum at several
tens of $R_{\rm e}$.  For the dark mass fraction within 1$R_{\rm e}$,
the most likely value is $\approx0.2-0.4$, comparable to the 
\citet{coccato2009} value ($\approx0.4$). The best fit stellar 
mass-to-light ratio $\Upsilon^B_\bigstar\sim 5\Upsilon^B_\odot$
is typical of metal-rich populations of age 5-6~Gyr \citep[see e.g.][]{bc03}.

\subsubsection{NGC~1023}
Our analysis favours a dark halo with $F_{\rm d}\approx9.9$ (near maximum)
and small radius $R_{\rm h}<10 R_{\rm e}$.  The poor halo fits for this galaxy
are perhaps unsurprising given the poor fits to the ``stars only''
basic model.  The stellar $\sigma_{\rm los}$ profile is more centrally
peaked than the (photometric) S\'ersic model.  It was observed
\citep{noordermeer2008,coccato2009} that the inner regions
$r<100\arcsec$) are affected by strong rotation. NGC~1023 is
a SB0 galaxy with a fast rotating bar which would imply 
a maximal (i.e. baryon dominated) disc \citep{vpd02}. We omit
NGC~1023 in the analysis.

\subsubsection{NGC~1344}
Fits to data from axis 2 are bad mainly because of the flat plateau in
velocity dispersion in the inner region. Axis~1 gives an acceptable
fit for $F_{\rm d}\sim$6, with a significant improvement over the
stars-only model.  According to \citet{teodorescu2005}, there is a
3.8$\times10^{11}$M$_\odot$ dark matter halo out to 3.5$R_{\rm e}$. A rough
estimate of the baryon fraction -- assuming 90\% of the stellar mass
is contained within 3.5$R_{\rm e}$ \citep{grah05} -- gives a baryon fraction
within this region of 0.29, in agreement with our estimates ($f_B\sim
0.3$).

\subsubsection{NGC~1400}
This lenticular galaxy is one of the two bright members of a nearby
group \citep[along with NGC~1407,][]{go93}. We extract the stellar
kinematics from \citet{proctor2009}.  We take their assumed distance,
but use the B magnitude and S\'ersic model from \citet{spolaor2008}.
Assuming a cosmic baryon fraction, none of the polytropic halo models
fits as well as ``stars only.''  Perhaps this is attributable to the
discrepancies between $\sigma_{\rm los}$ profiles from different
observations \citep[noted by][]{proctor2009}. We also notice that
NGC~1400 has a significant amount of rotation \citep[$v/\sigma\sim
0.5$ at $R_{\rm e}$,][]{bert94}, which could possibly explain its +0.3~dex
residual with respect to the Fundamental Plane \citep{ps96}.

\subsubsection{NGC~1407}

This E0 galaxy is the brightest member of a nearby group
\citep[including NGC~1400,][]{go93}. The stellar data were extracted
from \citet{proctor2009}, and we refer again to \citet{spolaor2008} for
the light profile.  The polytropic model with cosmic dark matter
fraction gives a good fit, with $F_{\rm d}\approx7$ and truncation
radius $\sim$15$R_{\rm e}$. The halo models with $F_{\rm d}\ga9$ have
significantly poorer fits than for low and medium effective degrees of
freedom.

\subsubsection{NGC~3377}

Stellar data for this flattened (E5) galaxy show a pronounced
``S-bend'' in the $\sigma_{\rm los}$ profile.  All the fits are
formally bad.  On the first axis, a halo with high $F_{\rm d}$ brings
a slight -- but perhaps insignificant -- improvement.  On the second
axis, a dark halo does not improve the fit at all, and the $F_{\rm
d}=9.5$ and 9.9 curves are worse than the others.  We suspect that
complicated orbital structure (especially the undulating projected
kinematics) renders straightforward spherical models inadequate and
inappliable to NGC~3377.  Details of this profile have been attributed
to the influence of a central massive object and discy dynamics
\citep{gebhardt2003,copin04}.  \citet{coccato2009} note a twist in the
PN kinematics. We omit NGC~3377 from consideration in our conclusions.

\subsubsection{NGC~3379}
This galaxy is another of the notable systems that \citet{romanowsky2003}
observed to have a radially declining $\sigma_{\rm los}$ profile.  For
the \citet{coccato2009} data on two axes, the dark halo does not improve
the fits over a S\'ersic-only model.  Dark matter is allowed if the
halo has a large radius and low central concentration.  Cases with
$F_{\rm d}\ge9$ are much worse fits than for lower $F_{\rm d}$.

For comparison with the split-axial results, we also fit the projected
velocity dispersion data representing circular rings, which are
extracted from figure~11 of \citet{douglas2007} \citep[who incorporated
stellar data from][]{statler1999}.  A purely stellar model fits
poorly: if the orbits are isotropic then $\chi^2_r\approx4.1$.  If
dark matter is present in the cosmic fraction, then it improves the
fitting somewhat: the best $\chi^2_r\approx$2 occurs for $F_{\rm
d}\approx7$ (see lower middle panel, Figure~\ref{fig.isotropic}).
Isotropic models with $F_{\rm d}\ge9$ fit unacceptably worse than
those with $F_{\rm d}\le8$.

The simple S\'ersic model of the stellar mass
   may be inadequate to improve upon these fits.
A bump in the stellar $\sigma_{\rm los}$ profile
  suggests the presence of kinematic substructure.

\subsubsection{NGC~3608}
Including a dark halo does not improve on stellar-only fits.  The
minima -- which have very high values of $\chi_r^2$ -- either imply a
very compact halo ($R\sim$1$R_{\rm e}$) with $F_{\rm d}>9$ or an
extended halo ($\sim 500$$R_{\rm e}$) with $F_{\rm d}=1$. We
tentatively blame this outcome on the centrally peaky
$\sigma_{\rm los}$ profile.
If the S\'ersic profile describes the stellar mass
distribution poorly, then the fitting routine compensates by seeking a
high central dark density.  \citet{coccato2009} remark that this
galaxy had few PN detections, and possible contamination from NGC~3607
nearby.  We omit NGC~3608 from further detailed examination.

\subsubsection{NGC~4374}

This E1 radio galaxy (Messier~84) is a member of the Virgo cluster.
Fits to this galaxy favour the presence of a dark halo,
with $F_{\rm d}\approx8$ and $R\approx$100 to 1000$R_{\rm e}$.
Unusually, for the second axis, the  $F_{\rm d}=9$
case improves on the basic ``stars only'' model
(although $F_{\rm d}\approx8$ is still the best fit).

\subsubsection{NGC~4494}
This is another of the intriguing cases where \citet{romanowsky2003}
found a declining velocity profile implying a dark matter halo with a
very low density \citep{napolitano2009}.  From data on both axes of
\citet{coccato2009}, the fitting characteristics resemble those of
NGC~3379: dark matter brings no improvement.  A halo is allowed if its
outer radius far exceeds the observed region.

\subsubsection{NGC~4564}
Models fitted to this E6 galaxy strongly prefer the presence of dark
matter, but with radius of only a few $R_{\rm e}$.  Perhaps this
surprising conclusion is due to the centrally peaky stellar
$\sigma_{\rm los}$ profile.  The mass and orbital distributions may
depart significantly from a S\'ersic sphere.  \citet{coccato2009}
indicate that there is a rotation curve resembling ``an S0 galaxy
rather than ... an elliptical galaxy''. The unphysically low values of M/L
and of the baryon fraction confirm the fit for this galaxy must invoke
non-spherical models. We omit NGC~4564 from the analysis.

\subsubsection{NGC~4697}
This flattened elliptical (E6) galaxy has especially numerous PN
measurements.  Its velocity dispersion declines in the outskirts, and
the degree of rotation appears to decline outside $R_{\rm e}$
\citep{mendez2001,mendez2008,sambhus2006,delorenzi2008a,coccato2009}.
Despite the non-sphericity, our fits to both axes favour the
presence of a dark halo with $F_{\rm d}\approx7$ or $8$ consistently.
The extrapolated outer radius is large, $R_{\rm h}\gg100R_{\rm e}$.  Thus the
characteristics of this galaxy are consistent with the other
apparently good fits.

\subsubsection{NGC~5128}
For the famous nearby galaxy NGC~5128 (Centaurus~A) the axis 1 data
strongly favour the presence of a dark halo with cosmically mean
composition.  The best fit indicates $F_{\rm d}\approx$6 and
$R\approx 25R_{\rm e}$. 
This is fewer dark thermal degrees of freedom than in
our other plausible fits.  The halo radius also seems more compact
than usual. The prominent dust lane implies
a recent merger, which might have left persistent departures from
spherical hydrostasis in the dark sector as well.  (Persistent dark
pulsations or rotation could possibly change the apparent $F_{\rm d}$
obtainable from inapplicably static kinematics.)  Unfortunately the
results from axis 2 are inconclusive due to insufficient data.

\subsubsection{NGC~5846}
NGC~5846 is the central galaxy of a nearby group, and it appears round
(morphological type E0-E1) and non-rotating \citep{coccato2009}.  We
should have expected decisive fits.  The S\'ersic parameters are
debatable: by fitting stellar and PN photometry, \citet{coccato2009}
find $n=12\pm2$, $R_{\rm e}=2903\pm192\arcsec$.  However
\citet{mahdavi2005b} fit the light profile with $n=3.95\pm0.05$ and
$R_{\rm e}=69.5\pm5.5\arcsec$.  We adopt the latter parameter pair
(high S\'{e}rsic indices, normally paired with very large half-light radii
are usually indicative of a fitting degeneracy).  Our fits to axis 1
marginally favour a dark halo with a small outer radius,
$R_{\rm h}\approx$ 7.8$R_{\rm e}$,
and $F_{\rm d}\approx 1$.  On axis 2, the S\'ersic-only
fit is similar to a stellar$+$dark matter model.

\subsection{Features of the dark matter profile}
\label{s.features}

In a Lane-Emden sphere unaffected by other mass components, the
quasi-entropy can take any positive real value, and there is a
one-to-one relation between the entropy and the natural outer
truncation radius of the halo.  We find that this is untrue for a halo
perturbed by a \citet{prugniel1997} stellar component.  For $F_{\rm
d}\ga6$ or so, we numerically find that there is a minimum $s_{\rm d}$
at which the halo fringe becomes infinitely diffuse, failing to
truncate.  For many models the cosmic baryon fraction is unattainable
beyond a particular range of the entropy (even if the halo is finite).

\begin{figure*}
\begin{center} 
\begin{tabular}{ccc}
	 \includegraphics[height=5cm]{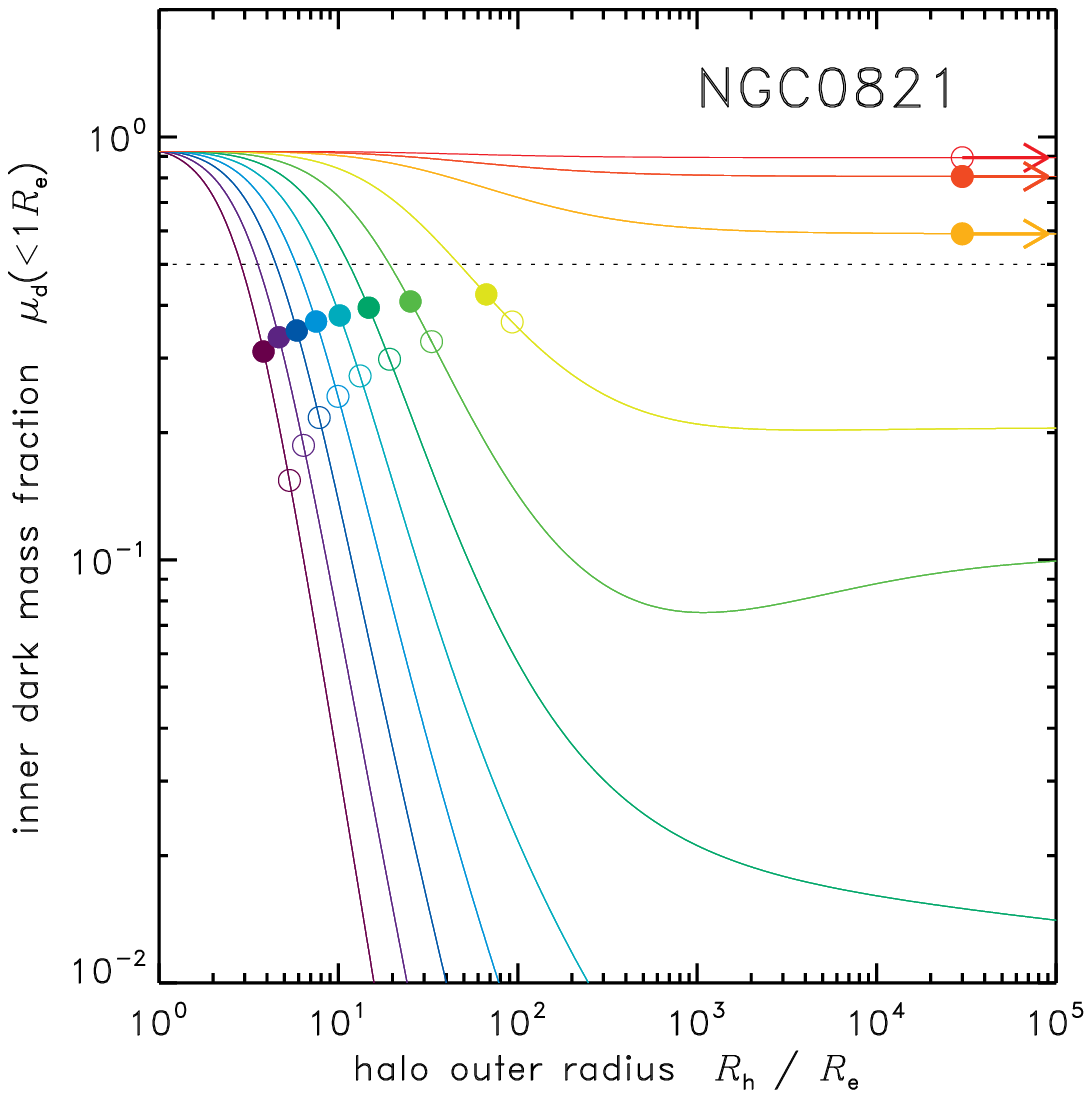}
	&\includegraphics[height=5cm]{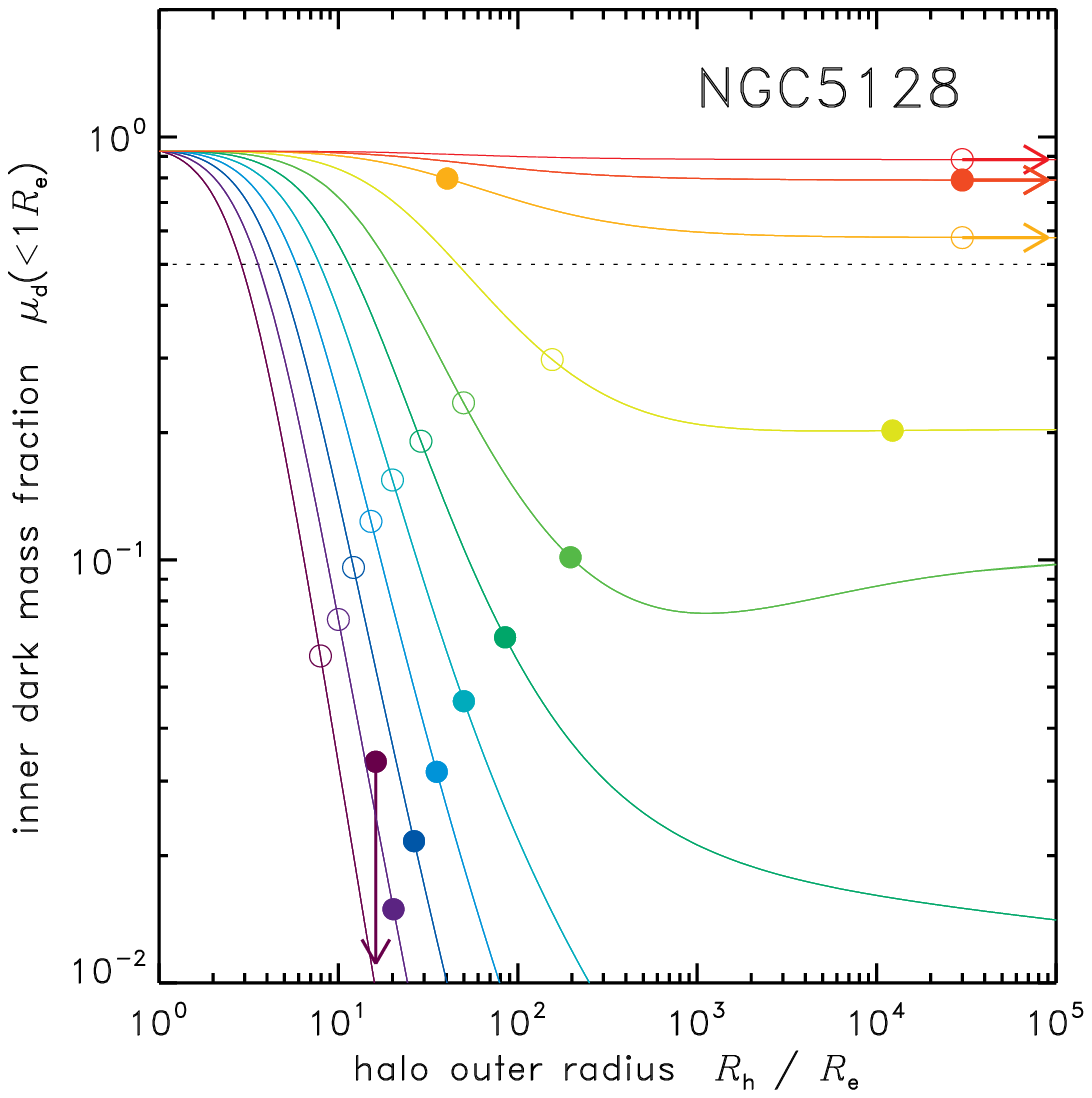}
	&\includegraphics[height=5cm]{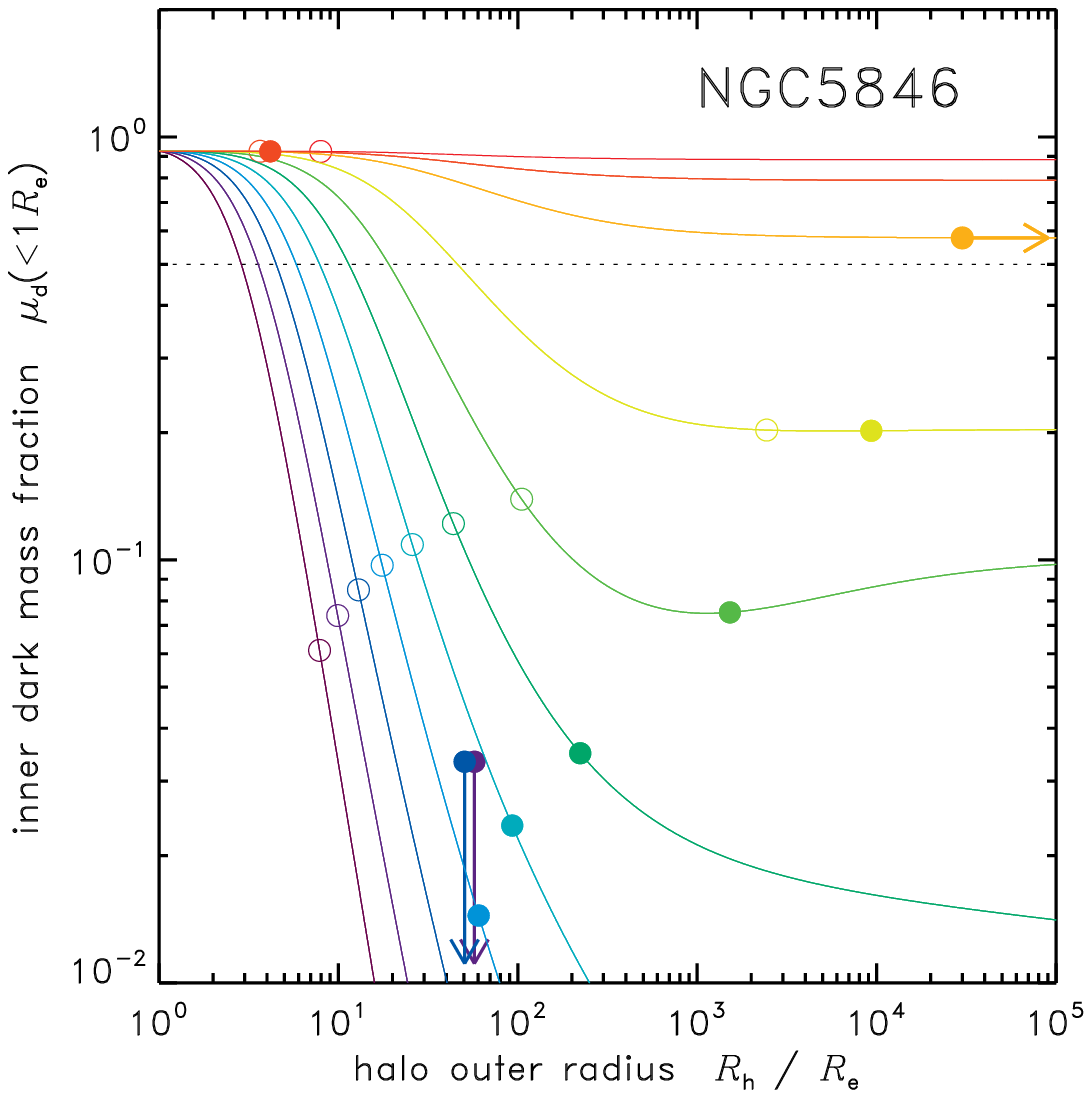}
\end{tabular}
\end{center} 
\caption{The dark mass fraction within 1$R_{\rm e}$, as function of
halo outer radius, for three galaxies with adequate fits.  Open and
closed circles mark the best fits to axis 1 and 2 respectively.  For
$F_{\rm d}=9, 9.5$ and 9.9 the dark core outweighs the local stars no
matter how far out the halo extends ($R_{\rm h}$). The horizontal
dotted line indicates the regions of baryon-dark matter equality.  }
\label{fig.dark.fraction}
\end{figure*}

The poor fitting by $F_{\rm d}\ga9$ halos has a related cause.  When
the degrees of freedom are numerous, the core is smaller and denser
compared to the outskirts.  This dense core tends to outweigh the
central stellar density, providing a poor fit to the inner data.  For
$F_{\rm d}\ga9$ the dark mass always outweighs the stars within
1$R_{\rm e}$ (see upper lines in figure~\ref{fig.dark.fraction}).  
Furthermore, when
$F_{\rm d}\ge9$ we find that inflating the halo (extending the fringe
radially) does not reduce the central dark density significantly.  For
lower $F_{\rm d}$, the core expands somewhat as the truncation radius
increases, and this enables fits with plausibly low dark densities
within the half-light radius.

\begin{figure*}
\begin{center} 
\begin{tabular}{ccc}
	 \includegraphics[width=8.cm]{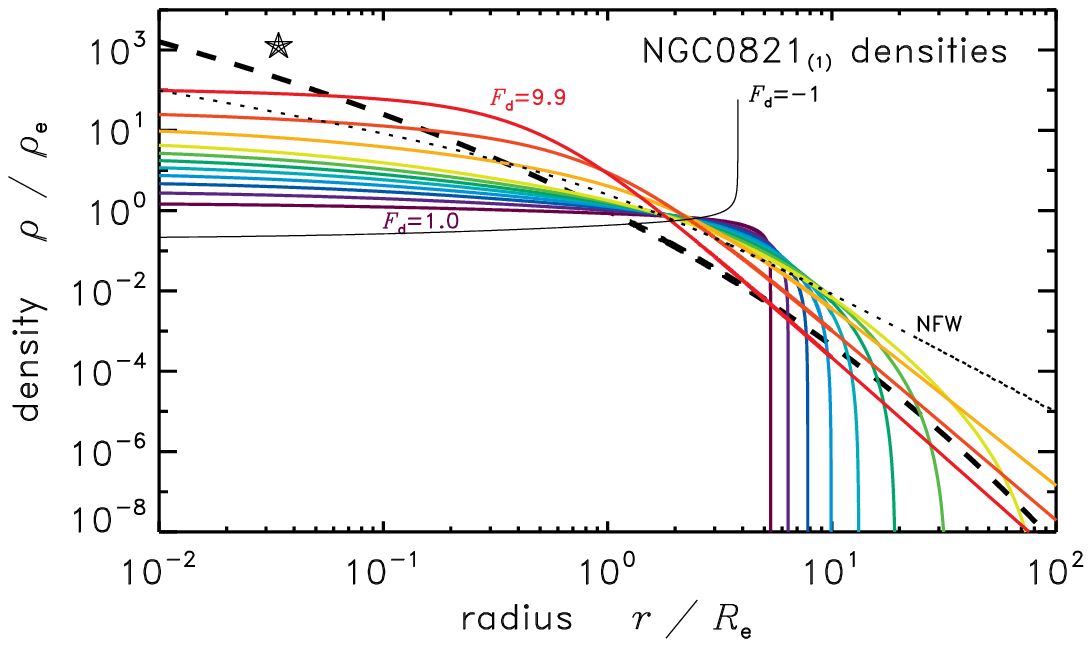}
	&\includegraphics[width=8.cm]{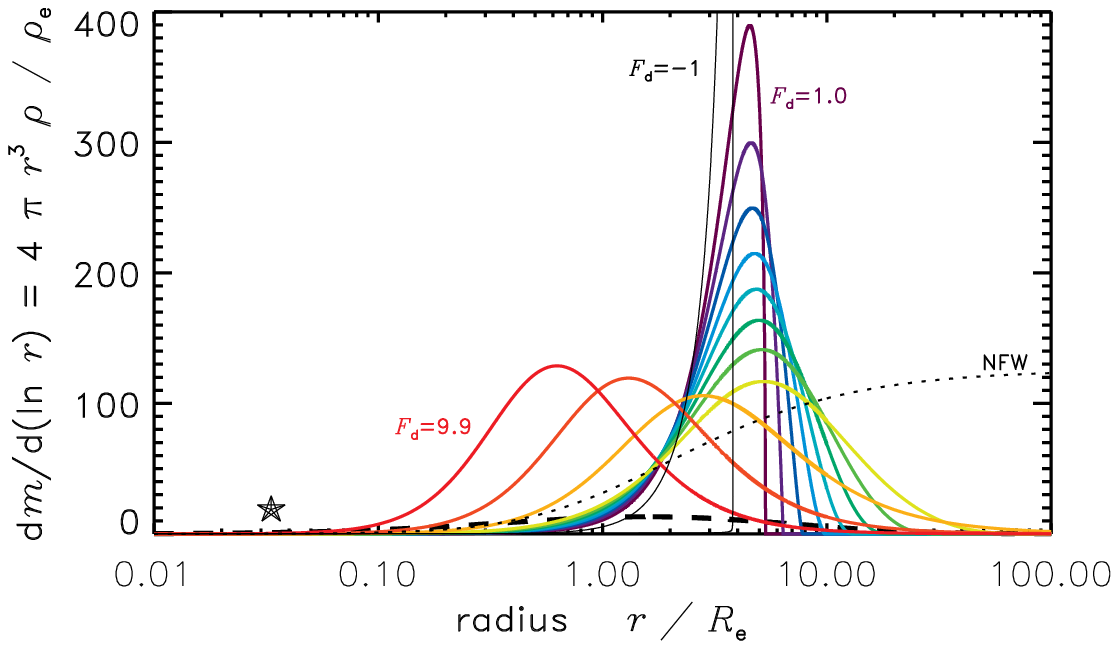}
\end{tabular}
\end{center} 
\caption{ The best-fit 3D halo density profiles for NGC~821 (axis 1)
  data.  The first panel shows the $\rho_{\rm d}(r)$ curves shaded as
  usual according to their $F_{\rm d}$ values, The dashed black
    line traces the stellar density profile for comparison.  The
    density normalisation $\rho_{\rm e}\equiv\rho_\bigstar(1R_{\rm
      e})$ is obtained in the optimisation of each fit.  (The total
    masses differ between $F_{\rm d}$ cases, but the stellar mass
    fraction is fixed.)  The dotted black line corresponds to a
    standard NFW profile \citep{nfw1996}.  The second panel shows a
    quantity $dm_{\rm d}/d(\ln r)$, which emphasises the layers where
    most of the dark mass resides.  Visually, the total mass is
    proportional to the area under each curve.  The $F_{\rm d}\ge9$
    fits concentrate much of the dark mass within the half-light
    radius, while $F_{\rm d}<9$ fits concentrate dark mass farther out
    than several $R_{\rm e}$.  The NFW model implies that most of the
    halo mass resides at the largest radii.  We also include a choice
    of negative $F_{\rm d}$ (thin line), which results in an
    increasing outward density (at decreasing temperature).}
\label{fig.fits.ngc0821}
\end{figure*}

Figure~\ref{fig.fits.ngc0821} illustrates the halo density profile for
the best-fitting models of axis 1 of NGC~821 for the choices of
$F_{\rm d}$ as shown in figure~\ref{fig.isotropic} (with the same
colour coding).  In our fits, this galaxy possesses a significant dark
mass fraction within the half-light radius, and the dark density at
1$R_{\rm e}$ is comparable to the stellar density there.  The
$F_{\rm d}\ge9$ curves are structurally distinct: the dark density
follows that of the stars closely for several decades in radius.  The
truncation radius is large, but nevertheless the central dark density
is high.  The more favourable fits, with $F_{\rm d}\le8$, involve
halos that truncate within 100$R_{\rm e}$, and there is a much larger
inner region where $\rho_{\rm d}>\rho_\bigstar$
(e.g. $r\approx$0.7$R_{\rm e}$ for $F_{\rm d}=8$).  For all fits, the
logarithmic slope of the dark matter halo density is around $-1$
around and within the half-light radius.  This should not be mistaken
for the $\rho_{\rm d}\propto r^{-1}$ cusps of collisionless N-body
simulations.  The innately flat core of the polytropic halo has been
gravitationally pinched by the stellar mass distribution.  The {\em
total} mass density index wavers around $-2$ out to several $R_{\rm e}$,
which may be consistent with implications from some gravitational
lensing studies \citep{rus03,koop06,if08}. If $F_{\rm d}<9$ and if our
models have typical mass concentrations then the polytropic halo of a
truly isolated elliptical galaxy would eventually be found to decline
more sharply than $\propto r^{-3}$, somewhere beyond
$\sim10R_{\rm e}$.  
Satellite galaxies and correlated cosmological macrostructure in
the vicinity may confuse the practical detection of this truncation.

\begin{figure*}
\includegraphics[width=8.5cm]{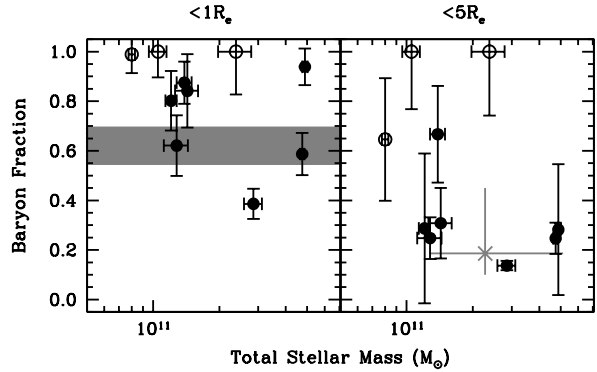}
\caption{The baryon fraction -- defined as
$m_\bigstar/(m_\bigstar+m_{\rm d})$ is shown within 1$R_{\rm e}$
({\sl left}) and 5$R_{\rm e}$
versus total stellar mass. We show the best fitting
cases that improve over the corresponding ``stars only''
models. Error bars are shown at the 68\% level. The solid dots
correspond to galaxies with a good fit $\chi_r^2<3$. For comparison,
the shaded area in the left panel is the result of \citet{bol08}
for gravitational lenses in the SLACS survey. The grey cross with
95\% error bars is lens B1104-181 from \citet{ferreras2005}, whose
lensing and stellar masses are obtained within 3.7$R_{\rm e}$.}
\label{fig.scatter_dark}
\end{figure*}

Figure~\ref{fig.scatter_dark} shows the scatter of the baryon fraction
with respect to stellar masses for the best-fit models that adequately
improve over the stellar-only baseline cases. Solid dots represent
those galaxies with a good fit ($\chi^2_r<3$).  The figure shows that
within 1$R_{\rm e}$ most galaxies appear baryon dominated except for
massive ones, mainly NGC~1407. At 5$R_{\rm e}$ the baryon fraction
decreases in most cases to $\sim$20\%. For comparison, the analysis of
strong gravitational lenses from SLACS \citep{bol08} is shown as a
grey shade in the left panel -- the SDSS-selected lenses mainly probe
the inner ($R\lesssim R_{\rm e}$) regions. There is also consistency with
the estimates of \citet{cappellari2006} using the SAURON sample --
they quote a dark matter fraction of $\sim 30\%$ in the inner
$R_{\rm e}$; or with the recent work of \citet{tor09}, with similar
values. The grey cross and error bars in figure~\ref{fig.scatter_dark}
({\sl right}) correspond to the lensing analysis of the galaxy
B1104-181 (at a redshift z=0.73) from \citet{ferreras2005}, which was
observed within $R<3.7R_{\rm e}$ (limited by the position of the lensed
images of the background source, which controls the accuracy of the
lensing estimates).

\subsection{Stellar brightness and mass fraction}

\begin{figure*}
\includegraphics[width=8.5cm]{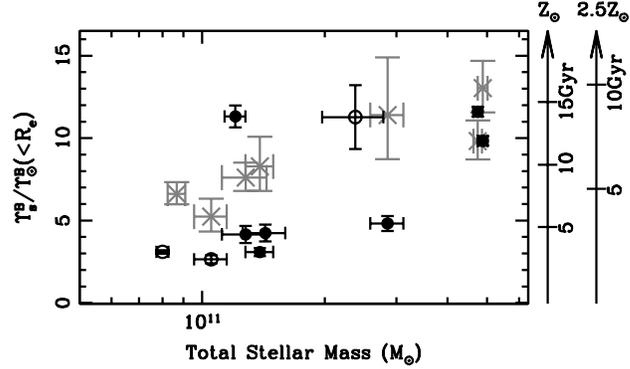}
\caption{The $B$-band stellar M/L ratio is shown versus total stellar
mass.  The points correspond to the best fitting cases that improve
over the corresponding ``stars only'' models (error bars shown at the
68\% confidence level). The solid dots are those galaxies with a
$\chi^2_r<3$. The grey crosses are M/L values from \citet{vdm07} for
those galaxies that overlap with our sample.  The arrows on the right
give the stellar M/L values for a Simple Stellar Population from the
CB07 population synthesis models \citep[e.g.][]{bruz07} for a \citet{chab03}
IMF. Two metallicities are considered, as labelled.}
\label{fig.scatter_star}
\end{figure*}

Though we have only fourteen galaxies (and not all of them with
satisfactory fits) it is interesting to consider their apparent
properties as a population.  Figure~\ref{fig.scatter_star} shows the
relation between $B$-band stellar mass-to-light ratio
($\Upsilon_\bigstar$) and the total stellar masses ($m_\bigstar$),
which are obtained jointly from the fitting process. Those galaxies
with a good fit ($\chi^2_r<3$) are shown as solid dots. In grey we
also show the {\sl dynamical} M/L values from \citet{vdm07} for the
galaxies in our sample. Those values are typically obtained within
$R_{\rm e}$, and agree rather well with our estimates using a polytropic
dark matter halo. For comparison with stellar populations, we also
show in the right side of figure~\ref{fig.scatter_star} the expected
stellar M/L for a couple of synthetic stellar populations from the
CB07 models \citep[e.g.][]{bruz07} for a \citet{chab03} IMF. Two
metallicities are considered, as labelled.
We find that the more massive galaxies tend to have older stellar
populations with ages compatible with more detailed work 
based on equivalent widths \citep[see e.g.][]{tm05,benyam}

\begin{figure*}
\begin{center} 
\begin{tabular}{ccc}
\includegraphics[width=5cm]{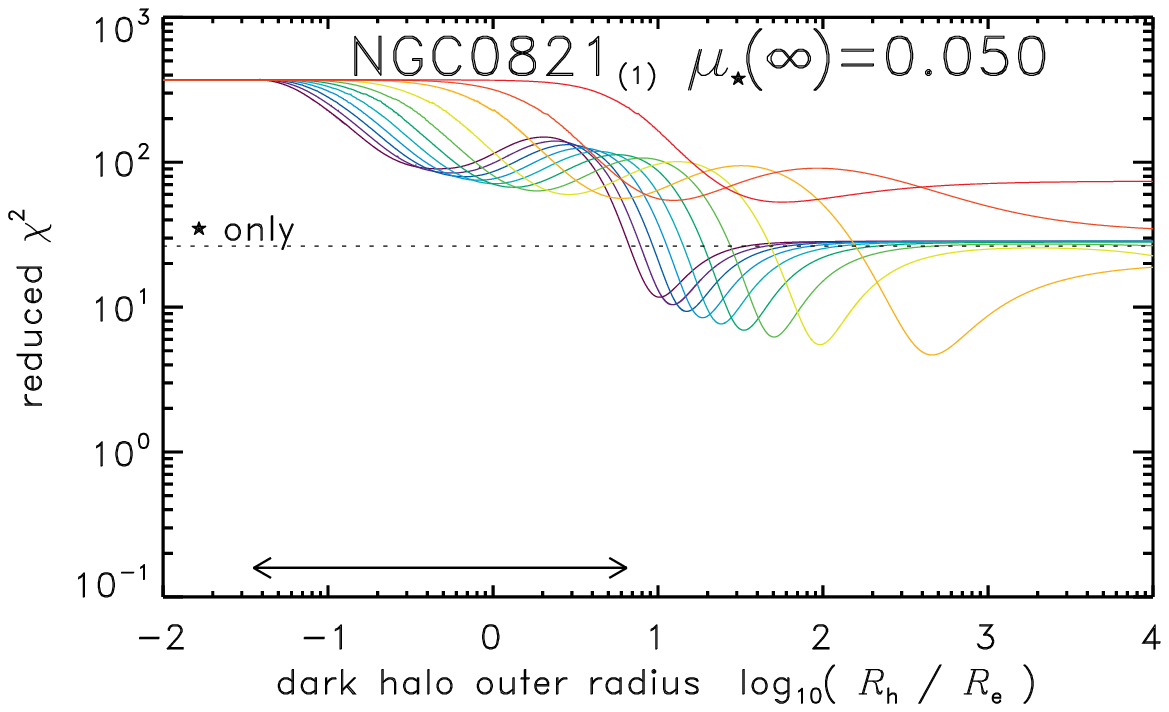}
&\includegraphics[width=5cm]{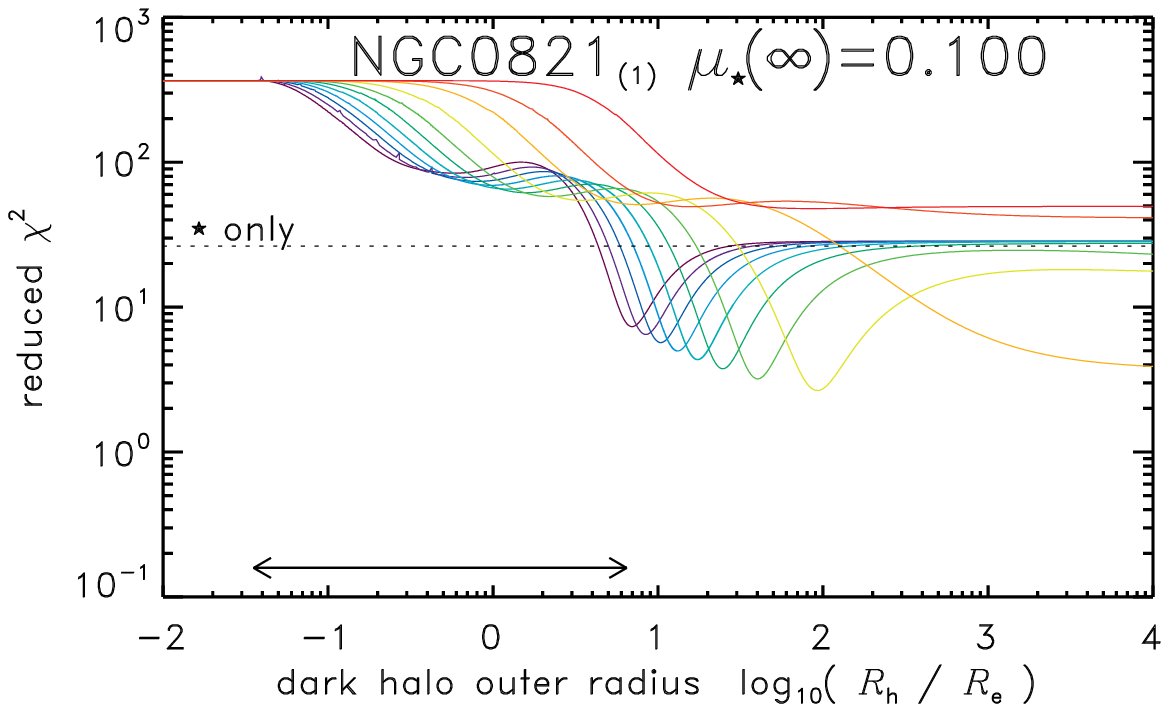}
&\includegraphics[width=5cm]{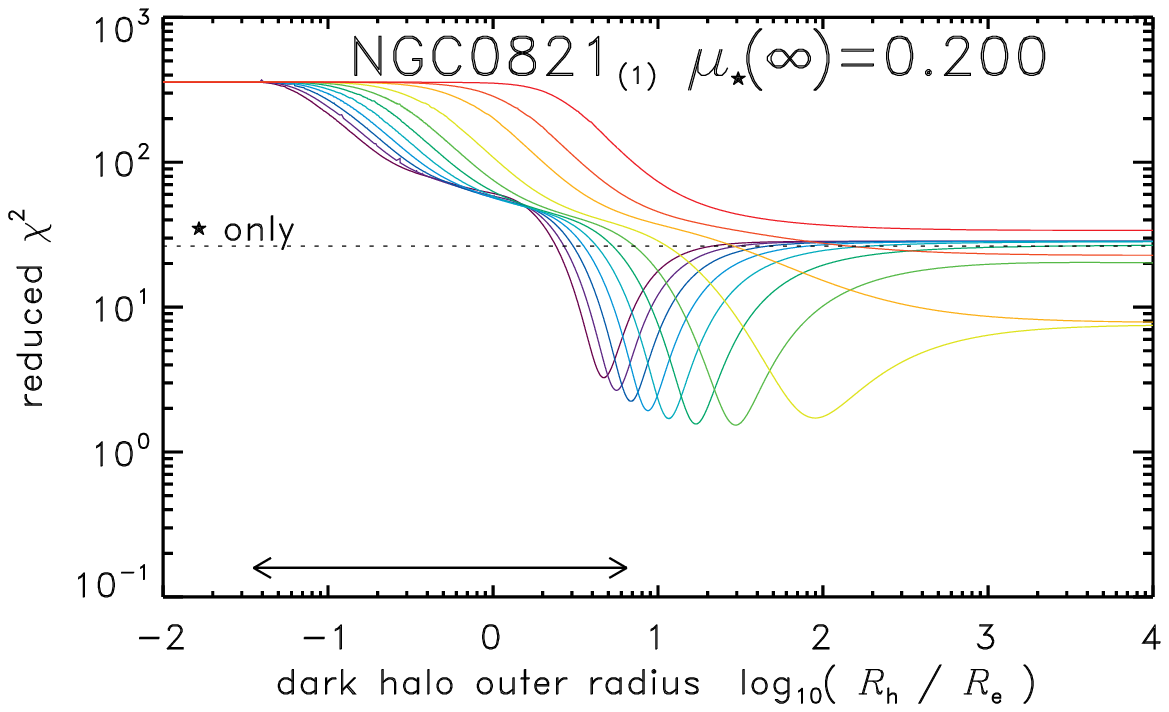}
\\
 \includegraphics[width=5cm]{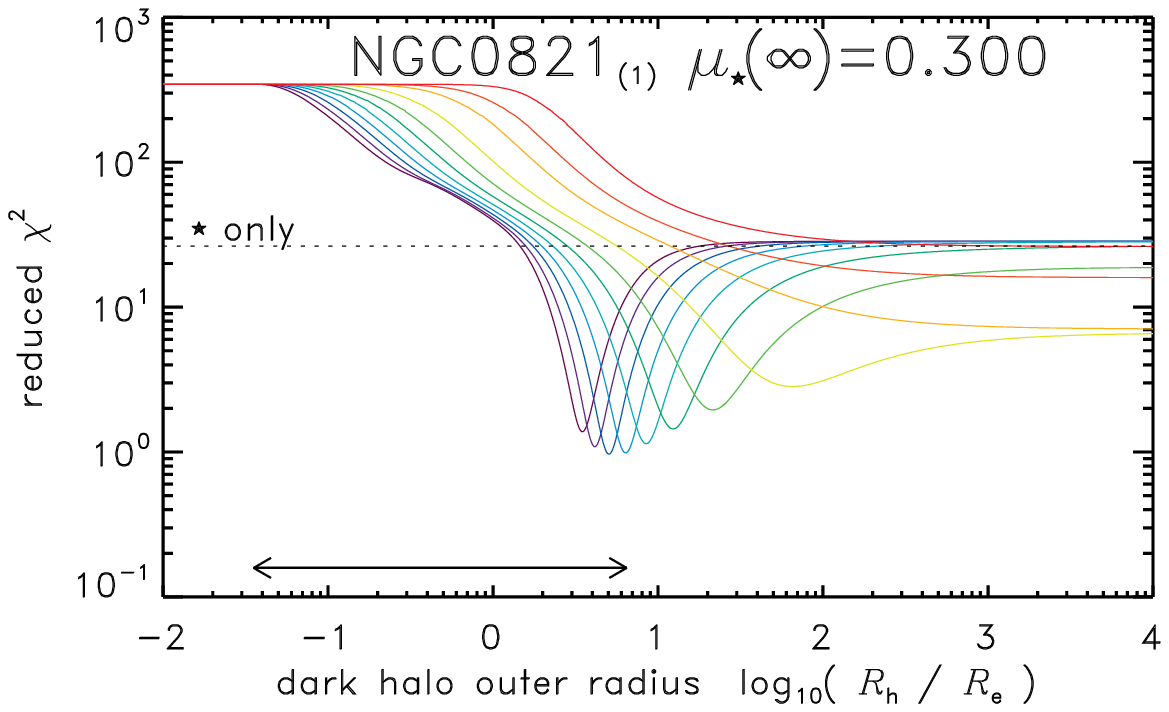}
&\includegraphics[width=5cm]{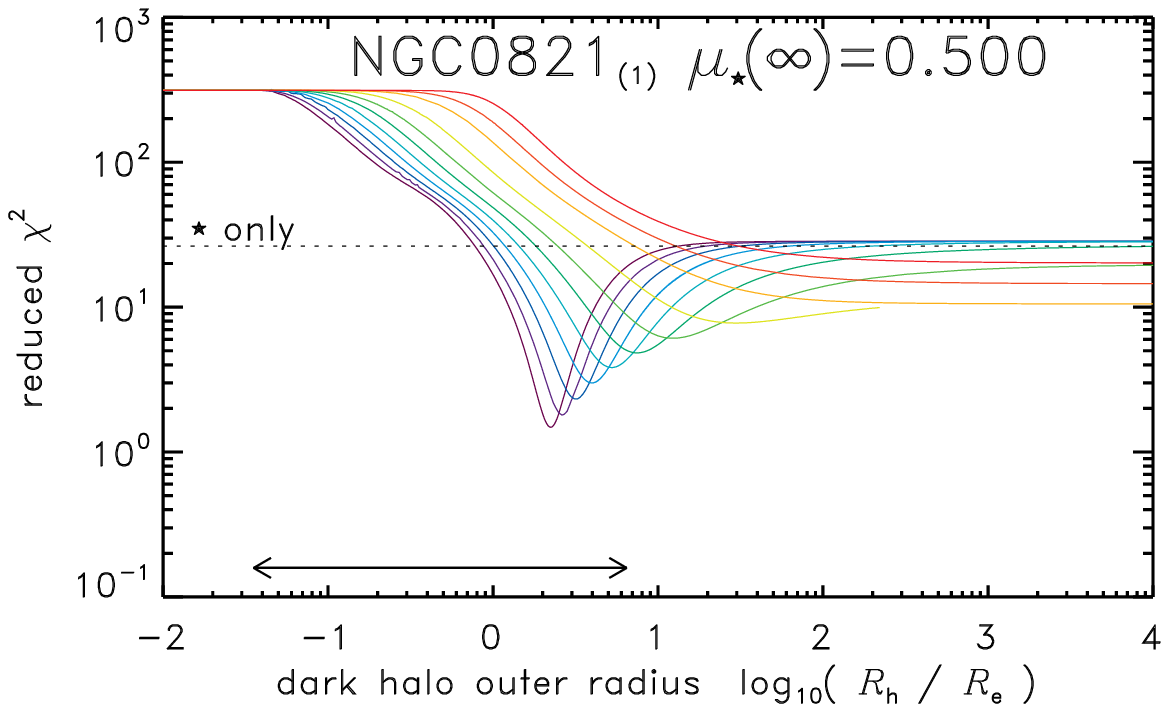}
&\includegraphics[width=5cm]{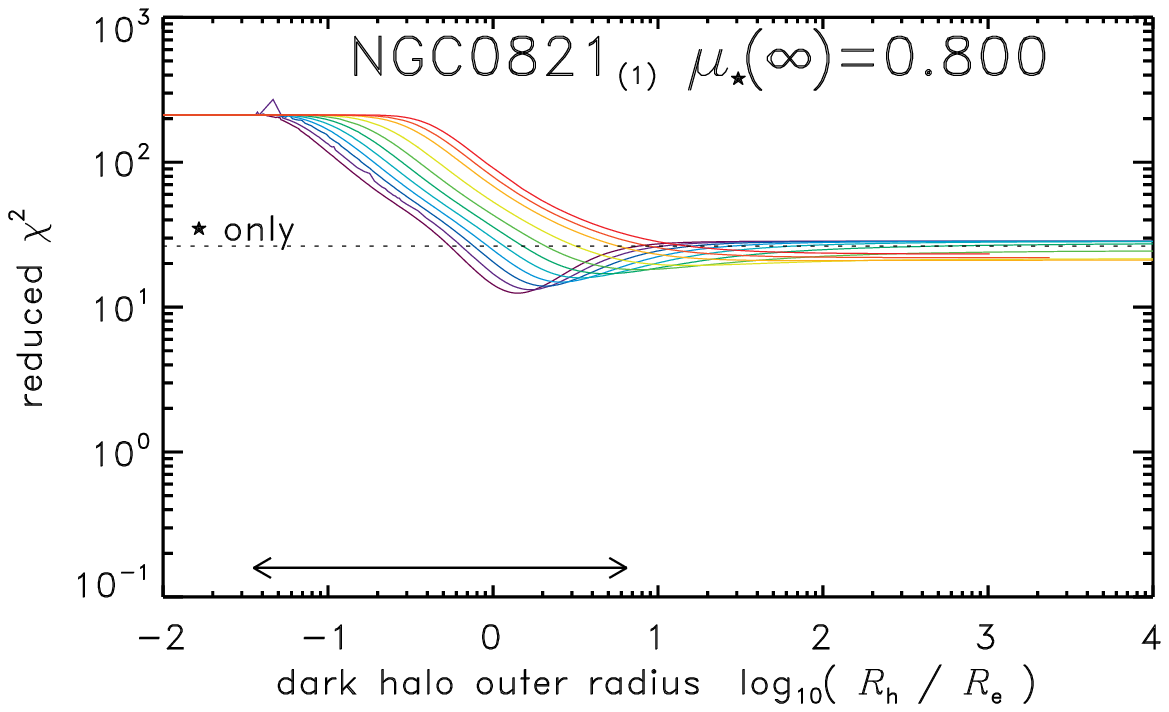}
\\
\end{tabular}
\end{center} 
\caption{ 
Reduced $\chi^2$ curves for halo models fitted to axis 1 of NGC~821,
   but with different stellar mass fractions, $\mu_\bigstar(\infty)$.
}
\label{fig.n0821.fstar}
\end{figure*}

Among our standard calculations, NGC~821 shows perhaps the most
convincing evidence for a dark halo of cosmic mass fraction.  In a set
of exploratory calculations, we vary the assumed ratio of stellar to
dark mass for this particular galaxy.  As illustrated in
Figure~\ref{fig.n0821.fstar}, the stellar mass fraction
$\mu_\bigstar(\infty)$ does have some effect.  If the galaxy is richer
in dark matter than the cosmic mean, then the best model shifts
towards slightly higher $F_{\rm d}$, e.g. for $\mu_\bigstar=0.05$ 
-- a value motivated by the latest estimates of the baryon fraction
in galaxy halos \citep[see e.g. ][]{moster09,guo09} -- the
optimum is around $F_{\rm d}\approx9$ with $R_{\rm h}\approx457R_{\rm e}$.  If
the galaxy is poor in dark matter then $F_{\rm d}$ and $R_{\rm h}$ both shift
to lower values.  As $\mu_\bigstar\rightarrow1$, the $\chi^2$ curves
collapse trivially to the poor fit of the ``stars only'' model.  The
best $\chi^2$ indicates moderate depletion of dark matter, e.g. when
fixing $\mu_\bigstar=0.30$ we find $\chi^2_r=0.98$ and
$R_{\rm h}\approx4.94R_{\rm e}$ for $F_{\rm d}=3$.  This would imply halo
truncation within the span of the observed planetary nebulae.  Models
with $\mu_\bigstar(\infty)\la0.2$ do not imply such a peculiarity.

\subsection{Effect of radial orbital anisotropy}
\label{s.anisotropy}

To explore the role of orbital anisotropy, we try re-fitting some of
the galaxies assuming the Osipkov-Merritt profile for $\beta(r)$ 
(see equation~\ref{eq.OM}).  In
this model, stellar orbits are isotropic in the centre, and
increasingly radial in the fringe beyond a scale radius $a$
\citep{osipkov1979,merritt1985}. For
``stars only'' fits the OM radial anisotropy profile provides
negligible improvement if the scale radius $a$ is large (approaching
isotropic models), and for small $a$ it makes the $\chi^2$ scores
significantly worse.  Therefore radial anisotropy in the outskirts
seems unlikely to cure the apparent dark deficit (or low halo
concentration) of cases like NGC~4494.  In this sense, ``stars alone''
models may be insufficient without non-spherical effects.

\begin{figure*}
\begin{center} 
\begin{tabular}{ccc}
\includegraphics[width=5cm]{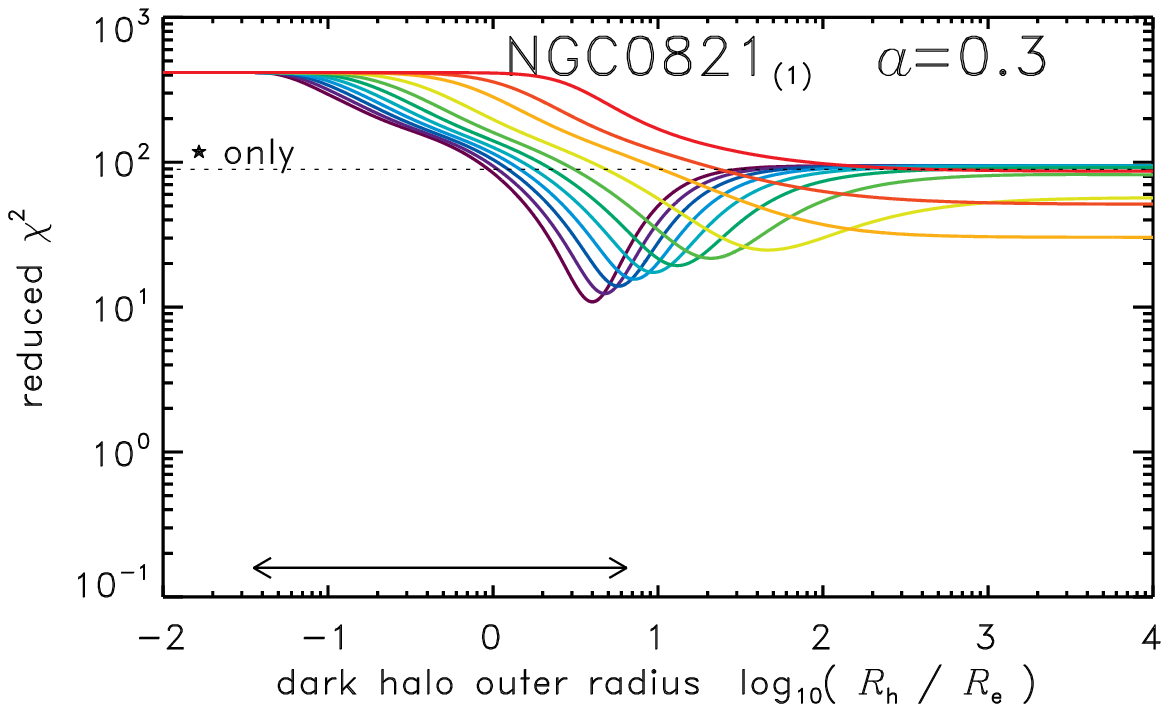}
&\includegraphics[width=5cm]{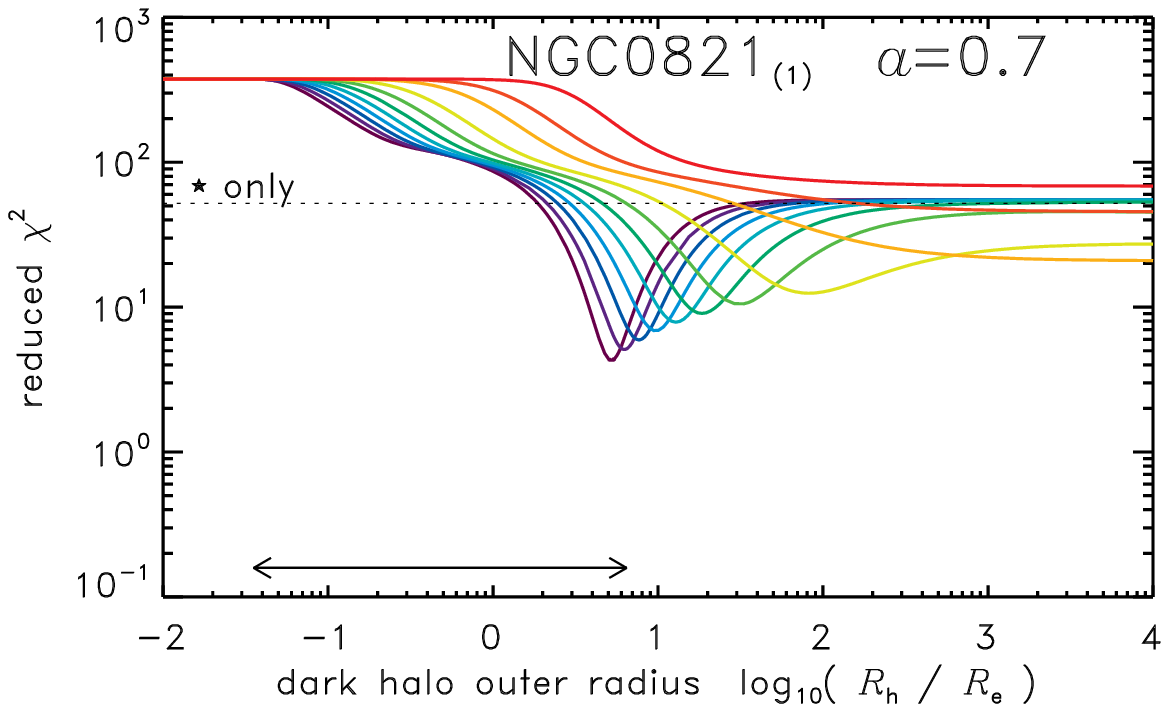}
&\includegraphics[width=5cm]{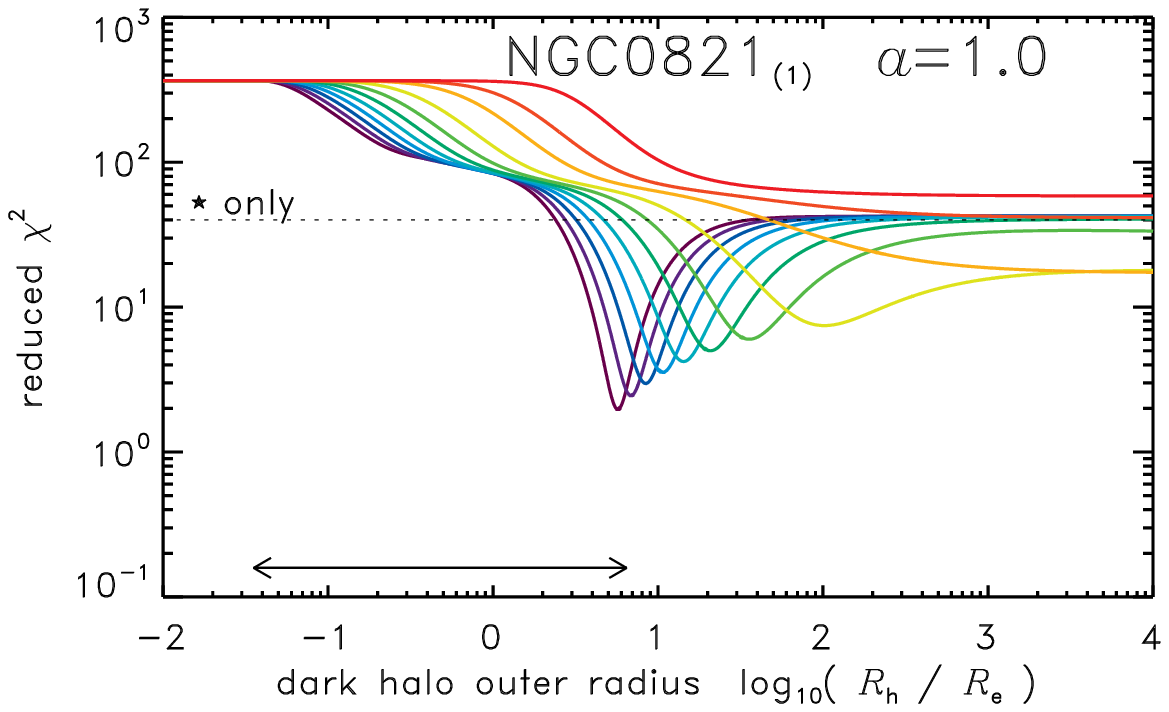}
\\
 \includegraphics[width=5cm]{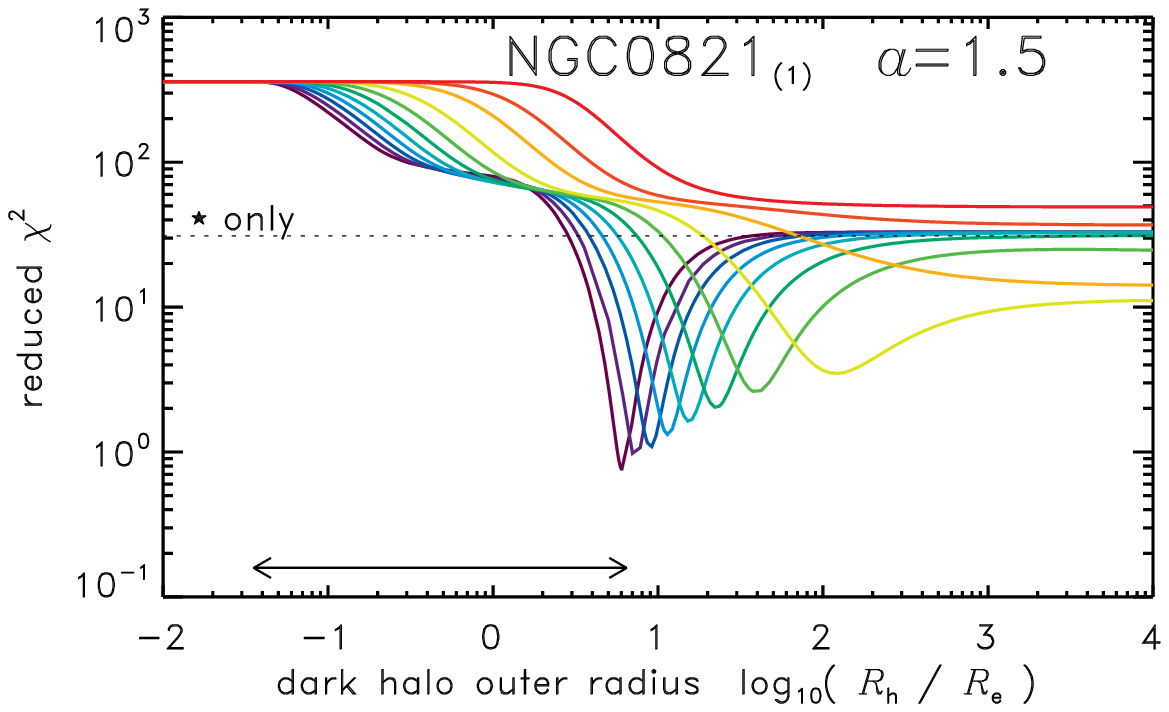}
&\includegraphics[width=5cm]{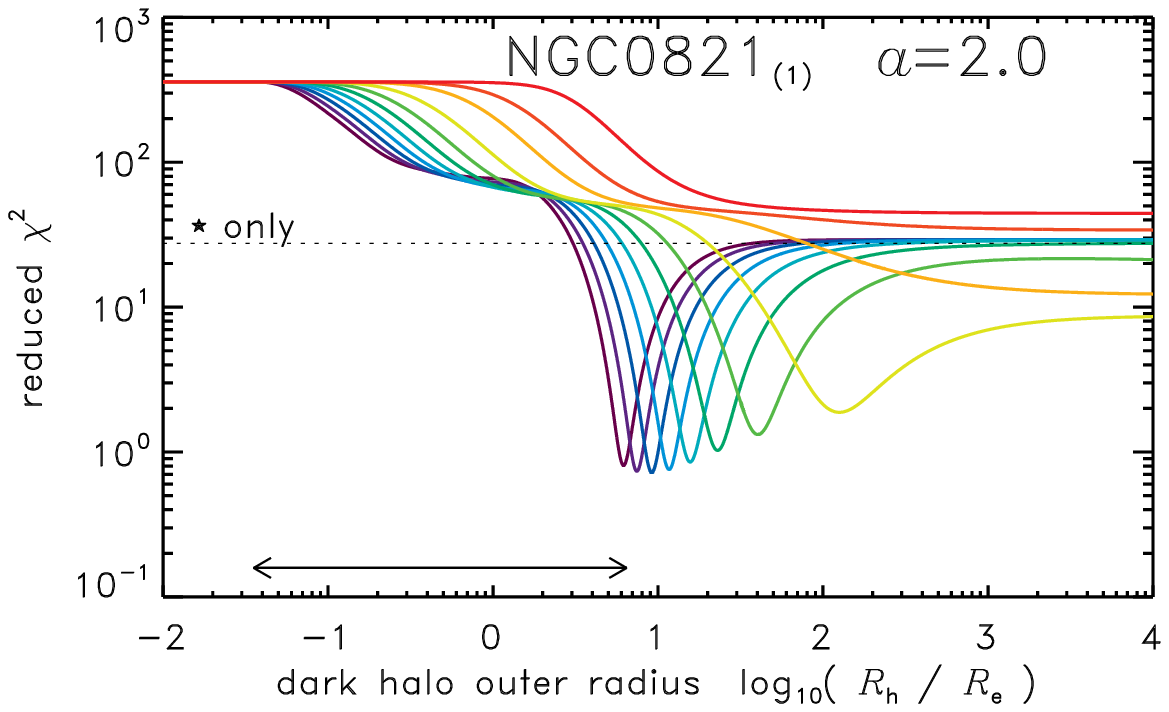}
&\includegraphics[width=5cm]{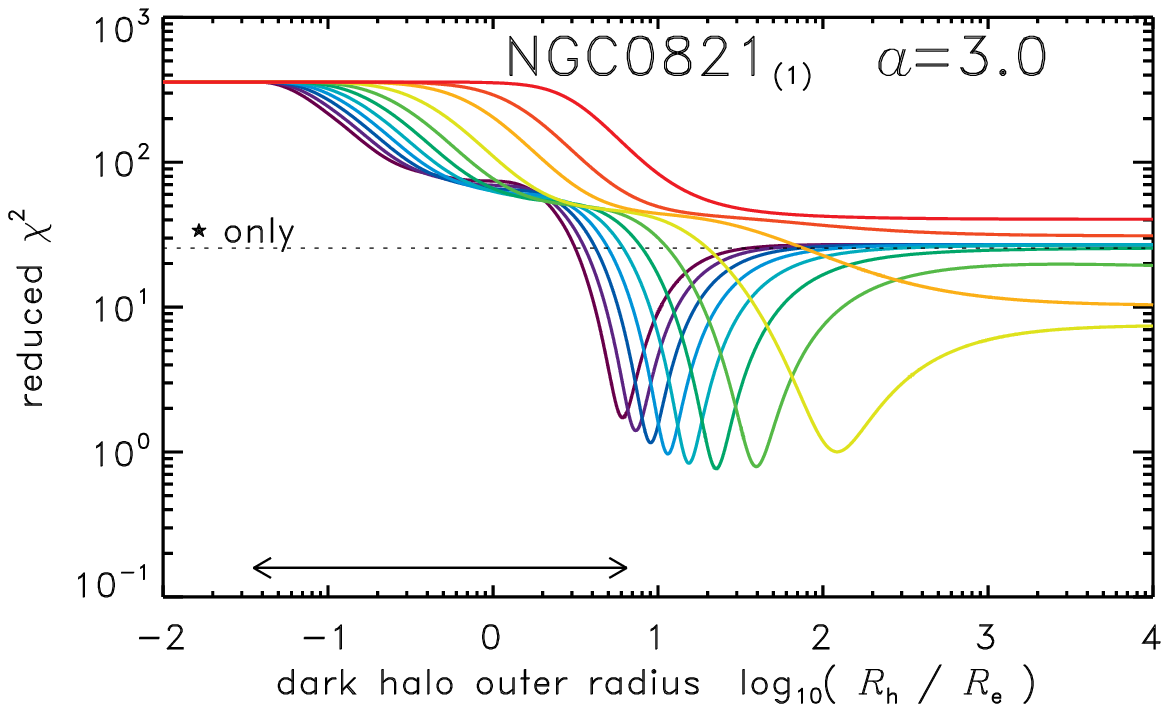}
\end{tabular}
\end{center} 
\caption{ 
Reduced $\chi^2$ scores of model fits to NGC821 axis 1,
   with curves of different $F_{\rm d}$ coloured as in
   Figure~\ref{fig.isotropic}.
Each panel shows a different choice of 
   the Osipkov - Merritt anisotropy scale radius $a$
   (in units of $R_{\rm e}$).
}
\label{fig.n0821.OM}
\end{figure*}

Figure~\ref{fig.n0821.OM} shows the effect of Osipkov-Merritt radial
orbital anisotropy profiles upon ``stars + halo'' models of NGC~821
(axis 1).  NGC~821 is one of the cases with a clear need for a dark
halo and this persists when the fringe is radially anisotropic.  The
$\chi^2$ curves still discriminate between $F_{\rm d}$ sharply.  The
curves prefer low $F_{\rm d}$ if $a<2$; for $a=2$ the optimum is
$F_{\rm d}=3$; for $a=3$ it is $F_{\rm d}=6$.  Thus the Osipkov-Merritt
radial anisotropy model favours small $F_{\rm d}$ if the scale radius
is comparable to the half-light radius, while isotropic models favour
$F_{\rm d}\approx8$.  The effects of the OM model are similar for
other galaxies, such as NGC~3379.  Overall $\chi^2$ levels are lower
for larger $a$, and thus the isotropic limit is formally favourable.
To convincingly constrain the amount and functional form of the
anisotropy profile, it would be necessary to fit more detailed models
with axisymmetry or fitting velocity kurtosis data.  Such complexity
is beyond the scope of our present paper.

\subsection{Exotic models: $-2<F_{\rm d}<0$}

It is worth mentioning the mathematical existence of
   a range of more exotic polytropes, with negative degrees of freedom.
Models with $-2<F_{\rm d}<0$ and negative pressure
   ($P_{\rm d}<0$, $s_{\rm d}<0$)
   describe a ``generalised Chaplygin gas,''
   which has been proposed to unify dark matter and dark energy
   \citep[e.g][]{bento2002,bilic2002,bertolami2004}.
Polytropes with this composition have infinite radius and mass,
   so they need arbitrary truncation at some background density
   \citep[e.g.][]{bertolami2005}.
Thus it is unclear how to naturally constrain the dark mass fraction
   in Chaplygin models of galaxies.

   However, if the pressure is positive, then the mass and radius of the
   halo are finite
   \citep{viala1974b,viala1974,kimura1981,lipscombe2008}.  In this
   ``anti-Chaplygin gas'' halo, density is minimal at the origin, and
   increases to infinity at the outer radius $R_{\rm h}$ where
   temperature drops to zero.  (Farther out, the dark density stays
   zero, as it does outside normal polytropes with $0<F_{\rm d}<10$.)
   These peculiar bubble-halo models are calculable, though the
   infinitely dense surface causes difficulty for the numerical
   integrators that project $\sigma_{\rm los}$.  We performed a sparse
   set of exploratory calculations in this regime, and find a
   continuation of the trends in $\chi^2(R_{\rm h},F_{\rm d})$ already
   seen at the low end of the $F_{\rm d}>0$ domain.  For NGC~821, the
   $\chi^2$ optimum rises (poorer fitting) and continues smoothly shifting to
   smaller $R_{\rm h}$ as $F_{\rm d}\searrow-2$.

   The limit of $F_{\rm d}=-2$ describes an isobaric condition.  In
   this case, hydrostasis requires the gravitational field to vanish
   everywhere.  There are no bound, finite, static, spherical,
   adiabatic solutions.  The choice of $F=-2$ and $P_{\rm d}=0$ could
   describe a collisionless medium, but the dark density and entropy
   $s_{\rm d}$ would then be arbitrarily spatially variable (which is
   a complication beyond the scope of this paper).

\section{Conclusions}

We have fitted published kinematics of fourteen diverse elliptical
galaxies, assuming a S\'ersic stellar mass profile plus a spherical
polytropic dark matter halo that is distributed self-consistently in
the shared gravitational field. This non-standard model is justified
by our current lack of knowledge about the nature of the dark matter
particle. A non-zero cross section for non-gravitational interactions
or internal degrees of freedom will justify this
interpretation. Furthermore, studies on the effect of gravitation on
an ensemble of particles suggest a polytropic equation of state is
required even if we are dealing with a non-interacting, elementary
particle \citep{tsallis1988}.

Our fits imply that more massive galaxies have higher stellar
mass-to-light ratios, and higher central fractions of dark matter.
We also impose a constraint on the degrees of freedom for the dark matter
particle, $F_{\rm d}\gtrsim 6$, in agreement with a previous study
focussed on galaxy clusters \citep{saxton2008}.
The favoured $F_{\rm d}$ increases to higher values if elliptical
galaxies are poorer in dark matter (and the favoured $F_{\rm d}$ is
smaller if baryons are lost).  Dark matter fractions well above or
below the cosmic fractions may be explicable in several ways.  Dark
matter may be lost via ablation or tidal stripping, especially in
dense environments. Baryonic matter might be lost due to blowout from
stellar or AGN feedback, or from inefficient cooling (at the massive
end), letting the baryons escape as warm or hot IGM/ICM unbound to
the galaxy.

The tentative preference for six to eight dark matter thermal degrees
of freedom is interesting when compared to circumstantial evidence
from galaxy clusters.  In their analyses of galaxy clusters comprising
polytropic dark matter and radiative gas, \citet{saxton2008} found that
the condition for stationary solutions implies a lower limit on the
central mass.  Compatibility with the scale of observed supermassive
black holes (as well as the dark matter core radii of clusters)
implies a constraint of $7\la F_{\rm d}\la10$.  (The upper limit is
necessary to form any finite self-gravitating body.)  Taken together,
the evidence from elliptical galaxies and galaxy clusters therefore
implies that dark matter has $F_{\rm d}\sim7$ to 8 (if it is the same
substance in both contexts).

Half of the available galaxies are well fit by the polytropic halo
model, despite its serious physical simplifications.  Some galaxies
may require tailored, non-spherical modelling or special treatment
of the kinematic relics of their ideosyncratic merger histories.  We
have assumed that the adiabatic dark halo is well mixed and is
non-rotating: this may be untrue for a galaxy that has merged less than
a few dynamical times ago.  Our present modelling precludes the
possible effects of a central star cluster or giant black hole, either
of which could induce a parsec-scale spike of dark and stellar matter
(within some gravitational radius of influence).  This could raise the
central velocity dispersion.  Allowing this feature to emerge in
future models might improve fits to the three galaxies with centrally
spiky $\sigma_{\rm los}(r)$ profiles.  

\section*{Acknowledgments}
We thank: K.~Wu for encouragement and criticism.
M.~Cappellari for a conversation prompting this work.
N.~Napolitano and L.~Coccato for their provision of online data.
\bibliographystyle{mn2e}
\bibliography{jour,blob,extra}


\begin{table*}
\begin{minipage}{16cm}
\caption{Reduced $\chi^2$ values for isotropic fits omitting PN data,
with a S\'ersic (stellar) mass component only.  There are too few data on axis 2
of NGC~5128, so in this case we fit both axes together. NGC~1400 and
1407 data \citep{proctor2009} are not split by axes. Primed values correspond
to a kinematic fit to the S\'ersic parameters.}
\label{table.stars.optima}
\begin{tabular}{lcccccccccccc}
\hline
NGC & $n$ & $R_{\rm e}$ & $\chi^2_{r,1}$ & $\chi^2_{r,2}$ &
$m_{\bigstar,1}$ & $m_{\bigstar,2}$ & $n^\prime$ & $R^\prime_{\rm e}$ & 
$(\chi^2_{r,1})^\prime$ & $(\chi^2_{r,2})^\prime$ &
$m^\prime_{\bigstar,1}$ & $m^\prime_{\bigstar,2}$\\
 & & ($\arcsec$) & & & \multicolumn{2}{c}{$(\times 10^{10}M_\odot)$} 
&  & ($\arcsec$) & & & \multicolumn{2}{c}{$(\times 10^{10}M_\odot)$}\\
\hline
 821 & 4.7 &  39,8  &  32.2  &  13.3  & 17.8 & 17.8  & 8.2  & 2310.  & 1.3 &  2.9 & 311. &  351.\\
1023 & 3.9 &  60.0  &   3.6  &   5.1  & 13.1 & 11.1  & 3.1  &   16.9 & 0.7 &  0.9 &   4.7 &   3.9\\
1344 & 4.1 &  50.0  &  84.2  &  22.2  & 15.0 & 12.5  & 8.3  &  765.1 & 8.2 & 14.7 &  70.5 &  68.0\\
1400$^1$ & 4.0 &  26.6  &   7.9  &   --   & 23.7 &  --   & 9.4  &  135.8 & 1.8 & --  &  43.9 & -- \\
1407$^{1,2}$ & 8.3 &  67.4  &  23.5  &   --   & 44.5 &  --   & -- & -- & -- & -- & -- & --\\
3377$^2$ & 5.2 &  54.0  &  34.6  &  19.6  &  0.2 &  0.3  & -- & -- & -- & -- & -- & --\\
3379 & 4.7 &  47.0  &  16.5  &   4.9  &  0.9 &  0.8  & 8.9  & 275.9 & 4.4 &  2.4 &  19.2 &  18.7\\
3608 & 7.0 & 157.0  &  50.1  &  49.8  & 31.0 & 30.0  & 3.5  &   8.9 & 6.2 &  7.0 &   4.3 &   4.2\\
4374 & 8.0 & 142.0  &   4.6  &   6.2  & 51.4 & 50.9  & 8.1  & 344.6 & 2.1 &  2.5 & 102.0 & 101.0\\
4494 & 3.3 &  49.0  &   7.4  &   7.6  & 10.5 & 11.0  & 7.2  & 490.5 & 5.5 &  2.1 &  41.2 &  39.0\\
4564 & 3.1 &  33.8  & 935.0  & 572.0  &  0.7 &  0.8  & 2.3  &   3.4 & 4.8 & 22.2 &   1.0 &   1.0\\
4697 & 3.5 &  66.0  &   0.4  &   2.3  & 14.5 & 14.5  & 5.8  & 614.3 & 0.3 &  1.2 &  79.5 &  76.1\\
5128$^2$ & 4.0 & 305.0  &   7.2  &   1.3  & 17.1 & 16.5  & -- & -- & -- & -- & -- & --\\
5846 & 4.0 &  69.5  &   7.6  &   5.8  & 51.3 & 51.2  & 8.1  & 1039.0 & 2.3 & 1.97 & 263.0 & 264.0\\
\hline
\end{tabular}
\end{minipage}
\\
$^1$ Only one axis available (data taken over a circular aperture).\\
$^2$ In these galaxies, the kinematic fit to the S\'ersic parameters does not converge.
\end{table*}


\begin{table*}
\begin{minipage}{16cm}
\caption{ Characteristics of the formally best fits for our stellar
$+$ polytropic dark matter model for each galaxy.  Columns from left
to right are: the galaxy identifier (subscripted with the axis -- axis
0 meaning circular annuli); the dark matter thermal degrees of
freedom; quality of fit for stars + halo ($\chi_r^2$); quality of fit
with stars alone ($\chi_{r,\bigstar}^2$; including PNe data);
truncation radius of the halo; stellar mass to light ratio in B band;
baryon fraction of the interior ($r<1R_{\rm e}$); baryon fraction of
middle regions ($r<5R_{\rm e}$). The uncertainties are quoted at the
1$\sigma$ level.}
\label{table.mol}
\begin{tabular}{lcccrrcrrr}
\hline
NGC & M$_B$ & M$_\bigstar\times 10^{10}M_\odot$ & $F_{\rm d}$ & $\chi_r^2$ & $\chi_{r,\bigstar}^2$ & 
$\log R_h/R_{\rm e}$ & $\Upsilon^B_\bigstar(\leq R_{\rm e})/\Upsilon^B_\odot$ & $f_B(\leq R_{\rm e})$ & 
$f_B(\leq 5R_{\rm e})$\\
\hline
 821$_{(0)}$ & --20.74 & $16.3\pm1.0$  &  7.9  &   1.9  &  3.0  & 6.00  &  $5.31\pm 0.32$  &   $0.80\pm 0.11$  &  $0.49\pm 0.26$ \\
 821$_{(1)}$ & --20.74 & $12.8\pm1.6$  &  8.0  &   1.7  & 26.4  & 1.97  &  $4.16\pm 0.52$  &   $0.62\pm 0.12$  &  $0.25\pm 0.08$\\
 821$_{(2)}$ & --20.74 & $13.6\pm2.7$  &  8.7  &   1.8  &  8.9  & 6.01  &  $4.42\pm 0.87$  &   $0.57\pm 0.23$  &  $0.23\pm 0.33$ \\
1023$_{(1)}$ & --20.92 & $ 0.4\pm4.9$  &  9.9  &   1.9  &  2.3  & 0.84  &  $0.12\pm 1.36$  &   $0.08\pm 0.39$  &  $0.14\pm 0.37$ \\
1023$_{(2)}$ & --20.92 & $ 0.3\pm4.9$  &  9.9  &   1.1  &  4.4  & 0.73  &  $0.08\pm 1.35$  &   $0.07\pm 0.45$  &  $0.14\pm 0.39$ \\
1344$_{(1)}$ & --19.59 & $12.1\pm0.7$  &  5.7  &   2.6  &  4.8  & 1.40  &  $11.32\pm 0.67$  &   $0.80\pm 0.12$  &  $0.29\pm 0.30$ \\
1344$_{(2)}$ & --19.59 & $13.8\pm1.8$  &  1.6  & 127.0  &111.7  & 8.88  &  $12.91\pm 1.69$  &   $>0.82$  &  $>0.71$ \\
1400$_{(0)}$ & --20.33 & $23.8\pm4.1$  &  0.2  &   6.5  & 6.1   & 2.74  &  $11.28\pm 1.93$  &   $>0.83$  &  $>0.74$ \\
1407$_{(0)}$ & --21.45 & $28.5\pm2.7$  &  7.0  &   0.3  & 19.8  & 1.18  &  $4.82\pm 0.45$  &   $0.39\pm 0.06$  &  $0.14\pm 0.02$ \\
3377$_{(1)}$ & --19.16 & $ 0.1\pm0.8$  &  9.9  &  22.1  & 25.1  & 0.32  &  $0.04\pm 1.11$  &   $0.08\pm 0.41$  &  $0.14\pm 0.38$ \\
3377$_{(2)}$ & --19.16 & $ 2.4\pm0.9$  &  0.1  &  17.5  & 17.3  & 0.63  &  $3.34\pm 1.31$  &   $0.87\pm 0.38$  &  $0.14\pm 0.38$ \\
3379$_{(0)}$ & --20.55 & $ 7.5\pm0.3$  &  6.8  &   1.95  & 3.25 & 1.96  &  $2.91\pm 0.13$  &   $0.88\pm 0.08$  &  $0.47\pm 0.26$ \\
3379$_{(1)}$ & --20.55 & $ 8.7\pm0.4$  &  1.7  &  13.8  & 11.1  & 9.46  &  $3.36\pm 0.17$  &   $>0.90$  &  $>0.73$ \\
3379$_{(2)}$ & --20.55 & $ 8.0\pm0.3$  &  0.4  &   3.3  &  3.17 & 1.07  &  $3.10\pm 0.11$  &   $>0.91$  &  $0.65\pm 0.25$ \\
3608$_{(1)}$ & --19.74 & $31.3\pm5.4$  &  1.0  &  50.54  & 42.8 & 2.69  &  $25.53\pm 4.37$  &   $>0.72$  &  $>0.64$ \\
3608$_{(2)}$ & --19.74 & $30.3\pm7.5$  &  9.9  &  44.37  & 42.6 & 0.19  &  $24.72\pm 6.14$  &   $>0.68$  &  $>0.63$ \\
4374$_{(1)}$ & --21.05 & $47.5\pm1.1$  &  8.6  &   0.1  &  2.8  & 3.13  &  $11.60\pm 0.28$  &   $0.59\pm 0.08$  &  $0.25\pm 0.06$ \\
4374$_{(2)}$ & --21.05 & $45.6\pm2.0$  &  8.6  &   0.4  &  4.4  & 2.17  &  $11.14\pm 0.49$  &   $0.44\pm 0.15$  &  $0.17\pm 0.14$ \\
4494$_{(1)}$ & --21.02 & $10.5\pm1.0$  &  1.6  &   3.9  &  4.0  & 8.88  &  $2.64\pm 0.24$  &   $>0.90$  &  $>0.77$ \\
4494$_{(2)}$ & --21.02 & $10.0\pm1.4$  &  1.6  &   5.9  &  5.8  & 8.88  &  $2.51\pm 0.36$  &   $>0.83$  &  $>0.70$ \\
4564$_{(1)}$ & -19.40 & $ 0.1\pm0.5$  &  9.9  &   3.17  & 625.8 & 0.64  &  $0.13\pm 0.01$  &   $0.07\pm 0.00$  &  $0.15\pm  0.05$ \\
4564$_{(2)}$ & -19.40 & $ 0.1\pm0.5$  &  5.0  &   0.62  & 370.5 & -0.74  &  $0.09\pm 0.01$  &   $0.07\pm 0.00$  &  $0.15\pm 0.06$ \\
4697$_{(1)}$ & --21.15 & $13.9\pm1.1$  &  7.4  &   0.2  &  0.4  & 6.23  &  $3.09\pm 0.24$  &   $0.87\pm 0.09$  &  $0.67\pm 0.19$ \\
4697$_{(2)}$ & --21.15 & $12.5\pm1.8$  &  8.1  &   0.5  &  1.2  & 6.00  &  $2.79\pm 0.39$  &   $0.75\pm 0.15$  &  $0.40\pm 0.29$ \\
5128$_{(1)}$ & --20.84 & $14.3\pm1.7$  &  5.7  &   0.6  &  7.1  & 1.41  &  $4.24\pm 0.51$  &   $0.84\pm 0.15$  &  $0.31\pm 0.14$ \\
5128$_{(2)}$ & --20.84 & $14.1\pm6.5$  &  0.1  &   1.0  &  1.3  & 1.12  &  $4.17\pm 1.93$  &   $>0.58$  &  $0.71\pm 0.37$ \\
5846$_{(1)}$ & --21.26 & $48.8\pm1.4$  &  0.9  &   1.0  &  2.3  & 0.89  &  $9.83\pm 0.28$  &   $0.94\pm 0.07$  &  $0.28\pm 0.26$ \\
5846$_{(2)}$ & --21.26 & $48.5\pm1.4$  &  6.5  &   0.8  &  0.7  & 5.94  &  $9.78\pm 0.29$  &   $0.95\pm 0.07$  &  $0.88\pm 0.24$ \\
\hline
\end{tabular}
\end{minipage}
\end{table*}


\label{lastpage}
\end{document}